\documentclass[traditabstract]{aa}
\usepackage{graphicx}
\usepackage{rotating}
\usepackage{natbib}
\bibpunct{(}{)}{;}{a}{}{,} 
\usepackage{dcolumn}

\newcolumntype{.}{D{.}{.}{-1}}
\newcolumntype{;}{D{;}{.}{7}}

\newcommand{\solm}{$M_{\odot}$\ }

\begin{document}

\authorrunning{Sabha, Eckart, Merritt, Zamaninasab et al.}  
\titlerunning{The S-star Cluster}
\title{The S-Star Cluster at the Center of the Milky Way \thanks{Based on observations 
collected at the European Organisation for Astronomical Research in the Southern 
Hemisphere, Chile (ProgId: 073.B-0085)}}
\subtitle{On the nature of diffuse NIR emission in the inner tenth of a parsec}
   \author{N.~Sabha\inst{1}
          \and
          A.~Eckart\inst{1,2}
          \and
          D. Merritt\inst{3}
          \and
          M.~Zamaninasab\inst{2}
          \and
          G.~Witzel\inst{1}
 	  \and
 	  M.~Garc\'{\i}a-Mar\'{\i}n\inst{1}
           \and 
 	  B.~Jalali\inst{1}
	  \and
 	  M.~Valencia-S.\inst{1,2}
 	  \and
 	  S.~Yazici\inst{1}
 	  \and
	  R.~Buchholz\inst{1}
 	  \and 
 	  B.~Shahzamanian\inst{1,2}
 	  \and 
 	  C.~Rauch\inst{1}
 	  \and
 	  M.~Horrobin\inst{1}
 	  \and 
 	  C.~Straubmeier\inst{1}
	  }

   \institute{ I.Physikalisches Institut, Universit\"at zu K\"oln,
              Z\"ulpicher Str.77, 50937 K\"oln, Germany\\
              \email{sabha@ph1.uni-koeln.de}
         \and
             Max-Planck-Institut f\"ur Radioastronomie, 
             Auf dem H\"ugel 69, 53121 Bonn, Germany
         \and
             Department of Physics and Center for Computational Relativity and Gravitation,
             Rochester Institute of Technology, Rochester, NY 14623, USA
             }

\date{Received: 9 March 2012/ Accepted:  4 June 2012 }


\abstract {Sagittarius~A*, the super-massive black hole at the center of 
the Milky Way, is surrounded by a small cluster of high velocity stars, known as the S-stars. 
We aim to constrain the amount and nature of stellar 
and dark mass associated with the cluster
in the immediate vicinity of Sagittarius~A*. 
We use near-infrared imaging to determine the $K_\mathrm{s}$-band luminosity function of the S-star 
cluster members, and the distribution of the diffuse background emission and the stellar number density counts around the central black hole. This allows us to determine the stellar light 
and mass contribution expected from the faint members 
of the cluster. We then use post-Newtonian N-body techniques to investigate the effect of
stellar perturbations on the motion of S2, as a means of detecting the number
and masses of the perturbers. We find that the stellar mass derived from the 
$K_\mathrm{s}$-band luminosity extrapolation is much smaller than the amount of mass that 
might be present considering the uncertainties in the orbital 
motion of the star S2. Also the amount of light from
the fainter S-cluster members is below the amount of residual light 
at the position of the S-star cluster after removing the bright 
cluster members. If the distribution of stars and stellar remnants is strongly enough peaked
near Sagittarius~A*, observed changes in the orbital elements of S2 can be used
to constrain both their masses and numbers. Based on simulations of the cluster of high velocity stars we find that
at a wavelength of 2.2~$\mu$m close to the confusion level for 8~m class telescopes
blend stars will occur (preferentially near the position of Sagittarius~A*) that last
for typically 3 years before they dissolve due to proper motions. 
}

\keywords{Galaxy: center - infrared: general - infrared: diffuse background - stars:  luminosity function, mass function - stars: kinematics and dynamics -  methods: numerical }

\titlerunning{The S-star Cluster}
\authorrunning{Sabha, Eckart, Merritt, Zamaninasab et al.} 
\maketitle

\section{Introduction}
\label{section:Introduction}

Using 8--10~m class telescopes, equipped with adaptive optics (AO) systems, at
near-infrared (NIR) wavelengths has allowed us to identify and study the 
closest stars in the vicinity of the super-massive black hole (SMBH) at 
the center of our Milky Way. These stars, referred to 
as the S-star cluster, are located within the innermost arcsecond, orbiting 
the SMBH, Sagittarius~A*(Sgr~A*), on highly eccentric and inclined orbits. Up till now, the 
trajectories of about 20 stars have been precisely determined using NIR imaging 
and spectroscopy \citep{ gillessen2009s2, gillessen2009stars}. This 
orbital information is used to determine the mass of the SMBH 
and can in principle be used to detect relativistic  effects and/or 
the mass distribution of the central stellar cluster 
\citep{rubilar2001,  zucker2006, mouawad2005, gillessen2009s2}. 

One of the brightest members of that cluster is the star S2. It has the shortest 
observed orbital period of $\sim$15.9~years, and 
was the star used to precisely determine  the enclosed dark 
mass, and infer the existence of a $\sim$4~million solar mass SMBH, in our own Galactic center 
\citep[GC;][]{schoedel2002, ghez2003}. The first spectroscopic studies 
of S2, by \cite{ghez2003} and later \cite{eisenhauer2005}, 
revealed its  rotational velocity to be that of  an O8-B0 young dwarf, 
with a mass of 15~\solm and an age of less than $10^6$~yrs. Later, 
\cite{martins2008} confined the spectral type of S2 to be a B0--2.5~V 
main-sequence star with a zero-age main-sequence (ZAMS) mass of 19.5~$M_\odot$.  
The fact that S2, along with most of the S-stars, is classified
as typical solar neighborhood B2--9~V stars, indicates that they are 
young, with ages between 6--400~Myr \citep{eisenhauer2005}.
The combination of their age and the proximity to Sgr~A*
presents a challenge to star formation theories. 
It is still unclear how the S-stars were formed. 
Being generated locally requires that their formation must have
occurred through non-standard processes, like formation in at least
one gaseous disk \citep{loeckmann2009} or via an eccentricity instability
of stellar disks around SMBHs \citep{madigan2009}. Alternatively, if they 
formed outside the central star cluster, about 0.3~parsec core radius
\citep[e.g.][]{buchholz2009, schoedel2007},
there are several models that describe how they may have been brought in 
\citep[e.g.][]{hansen2003, kim2004, levin2005, fujii2009, 
fujii2010, merritt2009, gould2003, perets2007, perets2009}. For a detailed 
description of these processes  
see \cite{perets2010}.
\begin{figure*}
  \centering
  \includegraphics[width=\textwidth,angle=00]{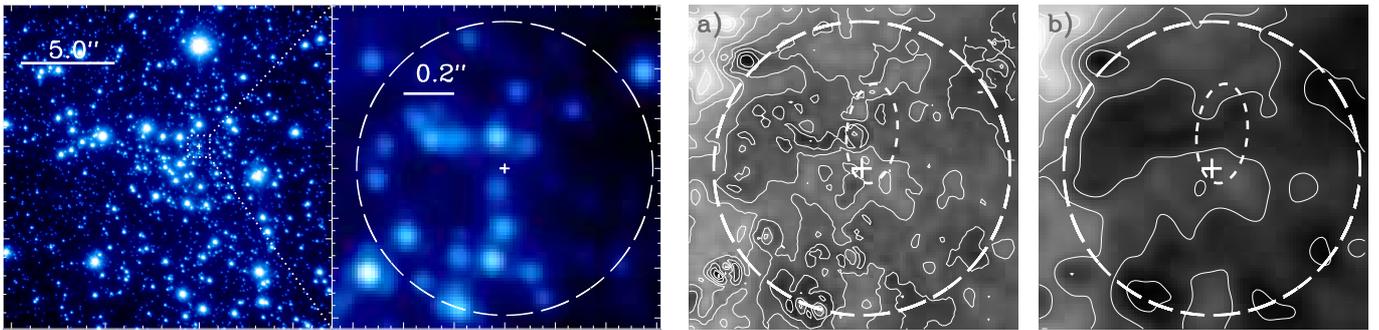}
  \caption{\small
  \textit{Left}: $17.5'' \times 17.5''$ NACO $K_\mathrm{s}$-band mosaic of the central cluster zoomed in to the 
  inner 1--$2''$ region around Sgr~A*. The inner region is indicated by a dashed (white) circle. 
  \textit{Righ}t: \textbf{a}) Map of the diffuse background light within a circle of $0.69''$ 
  radius centered on the position of Sgr~A*, shown here as a cross at the 
  center. The projected orbit of the star S2 is over-plotted as an ellipse. 
  \textbf{b}) The same map smoothed by convolution with a Gaussian beam of 
  \textit{FWHM}~$=6$~pixels. The contours levels are at 95\%, 90\%, 80\%, 70\%,
  60\%, 50\%, 40\%, 30\%, 20\% and 10\% of 
  the maximum flux value for each image.}
  \label{GCmap}    
\end{figure*}

Stellar dynamics predict the formation of a cusp of stars 
at the center of a relaxed stellar cluster around a SMBH. 
This is manifested by an increase in the three dimensional 
stellar density of old stars and remnants towards the center with 
power-law slopes of 1.5 to 1.75 
\citep{bahcall1976, murphy1991, lightman1977, alexander2009}. 

The steep power-law slope of 1.75 is reached in the case of a spherically 
symmetric single mass stellar distribution in equilibrium. For a cluster 
with differing mass composition, mass segregation sets in, where the more massive
stars sink towards the center, while the less massive ones remain less concentrated.
This leads to the shallow density distribution of 1.5 \citep{bahcall1977}. 
Later numerical simulations and analytical models confirmed these results 
\citep{freitag2006, preto2010, hopman2006}.
These steep density distributions were expected for the central cluster 
considering its age, which is comparable to the estimates of the two-body 
relaxation-time of 1--20~Gyr for the central parsec \citep{alexander2005, merritt2010, kocsis2011}. However, observations of the projected stellar number 
density, which can be related to the three dimensional density distribution, 
revealed that the cluster's radial profile can be fitted by two power-law slopes. 
The slope for the whole cluster outside a radius of $\sim 6''$ (corresponding 
to 0.22~parsec) was found to be as steep as $1.8\pm0.1$, while inside the 
break radius the slope was shallower than expected and reached an 
exponent of $1.3\pm0.1$ \citep{genzel2003stars, schoedel2007}.
These findings motivated the need to derive the density profiles of the distinct 
stellar populations, given that recent star formation \citep[6~Myr,][]{paumard2006}
at the GC gave birth to a large number of high-mass young stars that would be too 
young to reach an equilibrium state.
Using adaptive optics and intermediate-band 
spectrophotometry \cite{buchholz2009} found the distribution of 
late-type stars (K giants and later) to be very flat
and even showing a decline towards the Center (for a radius of less that $6''$), 
while the early-type stars (B2 main-sequence and earlier) follow a steeper 
profile. Similar results were obtained later by \cite{do2009stars} and \cite{bartko2010}.

These surprising findings required new models to explain 
the depletion in the number of late-type giants in the central few 
arcseconds around the SMBH. Such attempts involved Smooth Particle Hydrodynamics (SPH) 
and Monte Carlo simulations which tried to account for the under density of
giants by means of collisions with other stars and stellar remnants
\citep{dale2009, freitag2008}. Another explanation could be the disturbance
of the cusp of stars after experiencing a minor merger event or an
in-spiraling of an intermediate-mass black hole,
which then would lead to deviations from equilibrium; hence causing
a shallower power-law profile of the cusp \citep{baumgardt2006}.
\cite{merritt2010} explains the observations by the evolution of a parsec-scale 
initial core model.

\cite{mouawad2005} presented the first efforts to determine 
the amount of extended mass in the vicinity of the SMBH allowing for non-Keplerian orbits.
Using positional and radial velocity data of the star S2, and 
leaving the position of Sgr~A* as a free input parameter, they provide, for the first time, a rigid upper limit 
on the presence of a possible extended dark mass component around Sgr~A*.
Considering only the fraction of the cusp mass $M_\mathrm{S2_{apo}}$ that may be 
within the apo-center of the S2 orbit, \cite{mouawad2005}  find
$M_\mathrm{S2_{apo}}/(M_\mathrm{SMBH} + M_\mathrm{S2_{apo}}) \le 0.05$ as an upper limit. 
This number is consistent with more recent investigations of the 
problem \citep{gillessen2009stars}.
Due to mass segregation, a large extended mass in the immediate vicinity of Sgr~A*, if present, 
is unlikely to be dominated in mass of sub-solar mass constituents. 
It could well be explained by a cluster of high $M/L$ stellar remnants, 
which may form a stable configuration.

From the observational point of view, several attempts have been 
made recently to tackle the missing 
cusp problem. \cite{sazonov2011} 
proposed that the detected $1''$ sized thermal X-ray 
emission close to Sgr~A* \citep{baganoff2001, baganoff2003}
can be explained by the tidal spin-ups of several thousand late-type
main-sequence stars (MS). They use the \textit{Chandra} X-ray data to infer an upper 
limit on the density of these low-mass main-sequence stars. 
Furthermore, using \textit{Hubble} Space Telescope (HST) data, 
 \cite{yusefzadeh2012} derived a stellar mass 
profile, from the diffuse light profile
in the region $<1''$ around Sgr~A*,  and by that they explained the diffuse light to be dominated by 
a cusp of faint K0 dwarfs.

Up to now, the true distribution of the Nuclear Star Cluster, especially the S-stars, 
is yet to be determined. No 
investigations have confirmed or ruled out the existence 
of a cusp of relaxed stars and stellar remnants around Sgr~A*, 
as predicted by theory. 
An excellent dataset to investigate the stellar content of the central
arcsecond around Sgr~A* is the NIR $K_\mathrm{s}$-band (2.2~$\mu$m) data 
(see Figure~\ref{GCmap})
 we used in \citep[][hereafter NS10]{sabha2010}.
In that case we subtracted the stellar light
contribution to the flux density measured at the position of Sgr~A*. 
The aim of this work is then to analyze  the resulting image of the 
diffuse NIR background emission close to the SMBH. 
This emission is believed to trace the accumulative light of unresolved stars 
\citep{schoedel2007, yusefzadeh2012}. We explain the background light by extrapolating 
the $K_\mathrm{s}$-band luminosity function (KLF) of the innermost (1--2$''$, corresponding to 0.05~parsecs
for a distance of 8~kpc to the GC) members of the S-star cluster to fainter
$K_\mathrm{s}$-magnitudes. We compare the cumulative light and mass of these fainter stars to the 
limits imposed by observations. We then extend our analysis to explore the possible
nature of this background light by testing its effect on the observed orbit of the star S2.
Furthermore, we simulate the distribution of the unresolved faint stars 
($K_\mathrm{s} >18$) and their combined light to produce line-of-sight clusterings that
have a compact, close to stellar, appearance.

The paper is structured as follows: Section~\ref{Observations} deals with a brief description of the observation 
and data reduction. 
We describe in Section~\ref{section:CentralCluster}
the method used (\S~\ref{subsection:KLF}--\S~\ref{subsection:extraKLF}) and discuss 
the different observational limits (\S~\ref{subsection:Limits}) employed to test our 
analysis. Exploring the possible contributors to the dark mass within 
the orbit of S2 is done in Section \ref{section:DarkMass}. 
In Section~\ref{section:SimulateStars} we give the results obtained 
by simulating the distribution of faint stars and the
possibility of producing line of sight clusterings that look like
compact stellar objects.
We summarize and discuss the
implications of our results in Section \ref{section:Summary}.
We adopt throughout this paper $\Sigma(R) \propto R^{-\Gamma}$ as 
the definition for the 
projected density distribution of the background light 
, with $R$ being the projected 
radius and $\Gamma$ the corresponding power-law index.


\section{Observations and data reduction}
\label{Observations}
The observations and data reduction 
have been described in NS10.
In summary:
The near-infrared (NIR) observations 
have been conducted at the Very Large Telescope (VLT) of the 
European Southern Observatory (ESO) on Paranal, Chile. 
The data were obtained with YEPUN, 
using the adaptive optics (AO) module NAOS and the 
NIR camera/spectrometer CONICA (briefly ``NACO"). 
The data were taken in the  $K_\mathrm{s}$-band (2.2~$\mu$m) 
on the night of 23 September 2004,
and is one of the best available where Sgr~A* is in a quiet state. 
The flux densities were measured by aperture photometry with
circular apertures of 66~mas radius. They were corrected
for extinction, using $A_{K_\mathrm{s}} = 2.46$ 
derived for the inner arcsecond
from \cite{schoedel2010}.
Possible uncertainties in the extinction of a few tenths of a magnitude 
do not influence the general results obtained in this paper.
The flux density calibration was carried out using zero points 
for the corresponding camera setup and a comparison to known $K_\mathrm{s}$-band flux densities of IRS16C, IRS16NE (\citealp[from][]{schoedel2010};\citealp[ also][]{blum1996})
and to a number of the S-stars \citep{witzel2012}.



\section{The central few tenths of parsecs}
\label{section:CentralCluster}
In NS10 we gave a stringent upper  limit 
on the emission from the central black hole in the presence of the 
surrounding S-star cluster.  
For that purpose, three independent methods were used to remove 
or strongly suppress the flux density contributions of these stars, 
in the central $\sim2''$, in order to measure the flux density at 
the position of Sgr~A*. All three methods provided comparable 
results, and allowed a clear determination of the stellar light 
background at the center of the Milky Way, against which Sgr~A* 
has to be detected. The three methods, linear extraction of the extended flux density, 
automatic and iterative point spread function (PSF) subtraction 
were carried out assuming that the extracted PSF 
in the central few arcseconds of the image is uniform.
Investigations of larger images 
\citep[e.g.][]{buchholz2009} show that on scales of a few arcseconds 
the constant PSF assumption is valid, while for fields~$\ge 10''$ 
the PSF variations have to be taken into account.

\begin{figure}[!h]
  \centering
  \includegraphics[width=0.45\textwidth]{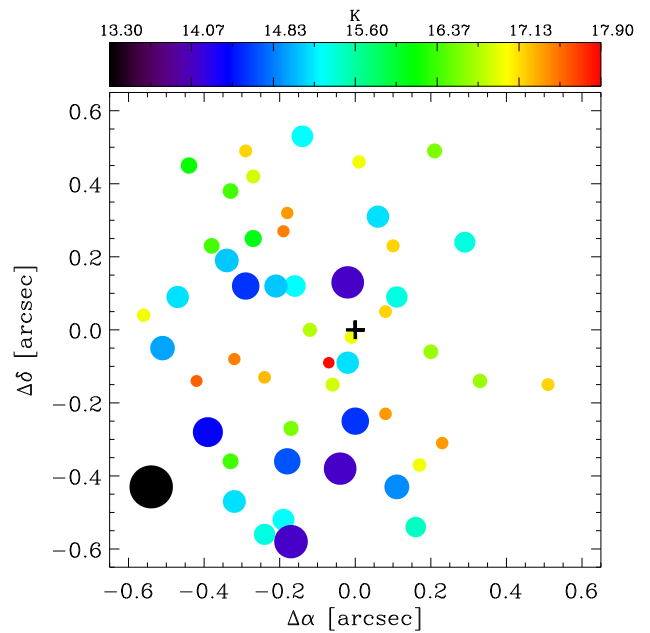}
  \caption{\small
  Map of the 51 stars listed in Table~3 from NS10. The color of 
  each star indicates its $K_\mathrm{s}$-magnitude. The size of each symbol is 
  proportional to the flux of the corresponding star. The position 
  of Sgr~A* is indicated as a cross at the center.
  }
  \label{starColor}    
\end{figure}

Figure~\ref{starColor} is a map of the 51 stars adopted 
from the list in Table~3 of NS10. The stars are plotted relative to the 
position of Sgr~A*. 
The surface number density of these detected stars, 
within a radial distance of about $0.5''$ from Sgr~A*, 
is $68\pm8$~arcsec$^{-2}$, with the uncertainty corresponding to the square-root 
of that value. This value agrees with the central number 
density of $60\pm10$~arcsec$^{-2}$ given by \cite{do2009stars}.
Extrapolating the KLF allows us to test if the observed 
diffuse light across the central S-star cluster, 
or the amount of unaccounted dark mass, can be 
explained by stars. 

\subsection{KLF of the S-star cluster}
\label{subsection:KLF}

\begin{figure}[!t]
  \centering
  \includegraphics[width=0.45\textwidth]{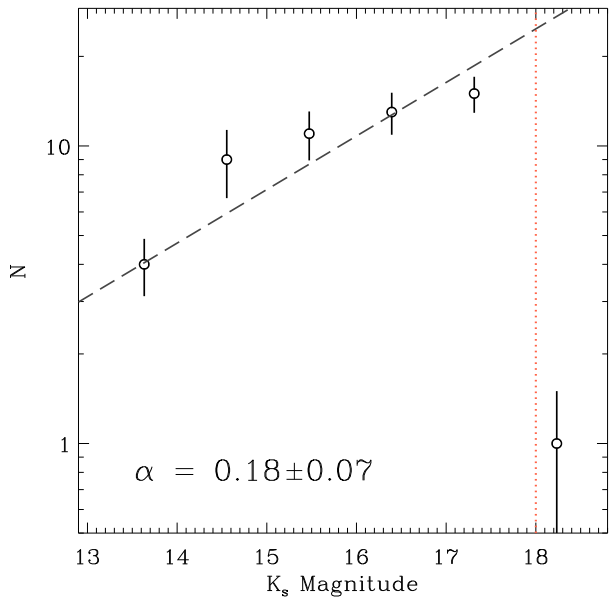}
  \caption{\small
  KLF histogram of the stars detected in the central field, 
derived from the 23 September 2004 data. 
The dashed line indicates the linear fit of the 
KLF slope of $ \alpha =0.18\pm0.07$. 
The vertical dotted line (red) represents the current 
detection limit for faint $K_\mathrm{s}$-magnitudes. 
  }
  \label{klf}    
\end{figure}

Figure~\ref{klf} shows the KLF histogram 
derived for the stars detected in the central field, (Figure~\ref{starColor}). 
We improve the KLF derivation by choosing a fixed number of 
bins that allows for about 10 sources per bin  
while providing a sufficient number of points 
to allow for a clear linear fit. The Red Clump (RC)/Horizontal Branch (HB) stars, 
around $K_\mathrm{s} \approx 14.5$, are in one bin, so the RC/HB bump is visible there \citep{schoedel2007}.
For estimating the uncertainty, we  randomized the start of the first 
bin in an interval between $K_\mathrm{s}=13.0$~to~$14.2$ 
and repeated the histogram calculation $10^5$ times. 
The number of sources in each bin was then determined by taking 
the average of all iterations and the uncertainties were subsequently 
derived from the standard deviation. We derive a least-square 
linear slope of $d\log(N)/d(K_\mathrm{s})= \alpha = 0.18\pm0.07$, 
which compares well with the KLF slope of $0.3\pm0.1$ derived 
in NS10 and also with the KLF slope of 
$0.21\pm0.02$ found for the inner field ($R<6''$) by \cite{buchholz2009}.
For the magnitudes up to $K_\mathrm{s}=17.50$ within the 
central $0.69''$ radius, we detect no significant 
deviation from a straight power-law. 
This implies that the completeness is high and can be 
compared to the $\sim$70\% value derived for 
mag$_K=17$ by \cite{schoedel2007} 
where the authors 
introduced artificial stars into their NIR image and attempted re-detecting them.
However, for $K_\mathrm{s}=17.50$~to~$18.25$ the stellar counts drop 
quickly to about $20\%$ of the value expected from the 
straight power-law line; hence the last $K_\mathrm{s}$-bin is excluded 
from the linear fit.
  
\cite{maiz2005} propose an alternative way of binning 
when dealing with stellar luminosity and initial mass 
functions (IMF). Their method is based on choosing variable 
sized bins with a constant number of stars in each bin. 
They find that variable sized binning introduces bias-free 
estimations that are independent from the number of stars per bin. 
Their method is applicable to small samples of stars. 
We apply their method to our KLF calculation and get  
$d(\log(N)/\delta K_\mathrm{s})/d(K_\mathrm{s})=0.12\pm0.09$, 
consistent with our fixed sized binning method.


\subsection{The diffuse NIR background}
\label{subsection:DiffuseLight}
The methods we used in NS10 to correct for the flux density 
contribution of the stars in the central $2''$ have revealed 
a faint extended emission around Sgr~A* (NS10 Figures~3b, 4b and 5). 
We detected $\sim1.3$~mJy
(obtained by correcting the $\sim2$~mJy 
we quote in NS10 for the $A_{K_\mathrm{s}} = 2.46$ we use here)
at the center of the S-star cluster. 
With a radius of $1''$ (about twice the FWHM of the S-star cluster) 
for the Point Spread Function (PSF) used for the 
subtraction, we showed that a misplacement of the PSF for 
about only five stars, located within one FWHM of Sgr~A*, 
would contribute significantly to the measured flux at the center. 
For a median brightness of about 
1.3~mJy
for these stars, 
a 1~pixel~$\sim 13$~mas positional shift of each of these stars 
towards Sgr~A* would be required to explain all the detected 
$\sim1.3$~mJy
at the center i.e. 
0.26~mJy
from each star. 
In \cite{sabha2011} we showed that a displacement larger than a 
few tenths of a pixel would result in a clear and identifiable 
characteristic plus/minus pattern in the residual flux distribution 
along the shift direction. For a maximum positional uncertainty of 1 pixel, 
we showed that the independent shifts of the five stars can be 
approximated by a single star experiencing five shifts in a 
random walk pattern. 
This resulted in calculating a total maximum contribution 
of 
0.26~mJy
from all the five stars to the center, which translates 
to about 20--30\% of the flux density. 
Thus, more than two thirds of the extended emission detected 
towards Sgr~A* could be due to faint stars, at or beyond 
the completeness limit reached in the KLF, 
and associated with the $\sim 0.5$--$1''$ diameter S-star cluster.

\begin{figure}[!h]
  \centering
  \includegraphics[width=0.45\textwidth]{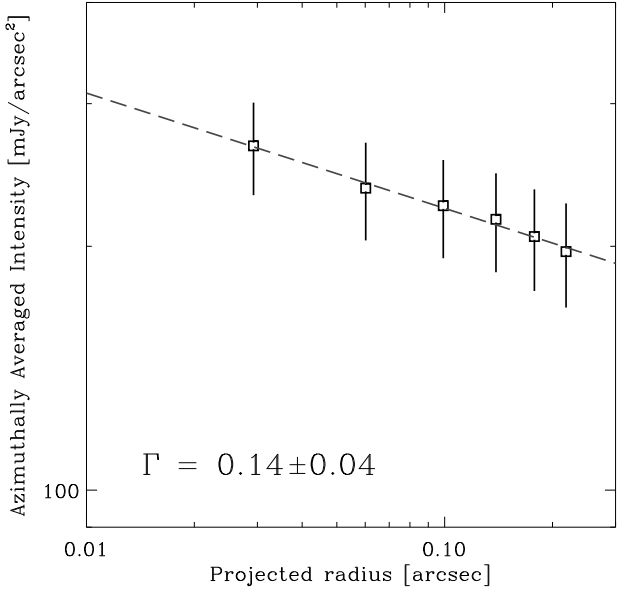}
  \caption{\small
  Azimuthal average of the diffuse background emission as derived 
  from manual PSF subtracted 23 September 2004 image.
  The squares (mean flux and $1 \sigma$ uncertainty per pixel) have 
  been calculated in annuli of 39.8~mas (3 pixels) width. 
  The black dashed line marks 
  a fit to the data points with an exponential decrease of 0.14.
  }
  \label{intensity}    
\end{figure}

The diffuse background emission we detected (see Figure~\ref{GCmap}a) could be compared to 
the projected distribution of stars $\Sigma(R) \propto R^{-\Gamma}$, 
with $R$ being the projected radius. 
We found that the distribution of the azimuthally averaged 
residual diffuse background emission, centered on the position 
of Sgr~A*, not to be uniform but in fact decreases gently as 
a function of radius (see Figure~7 in NS10) with a power-law 
index $\Gamma_{\mathrm{diffuse}}=0.20\pm0.05$. 
In this investigation we re-calculate the azimuthally averaged 
background light from the iterative PSF subtracted image alone.
The azimuthally averaged background light 
is plotted as a function of projected radius from Sgr~A* 
in Figure~\ref{intensity}. 
In this new calculation we find the power-law index to have a value of 
$\Gamma_{\mathrm{diffuse}}=0.14\pm0.07$. 
Both results are consistent with recent investigations 
concerning the distribution of number density counts of 
the stellar populations in the central arcseconds, derived 
from imaging VLT and Keck data. 
For the central few arcseconds
\cite{buchholz2009}, \cite{do2009stars} and \cite{bartko2010}
find a $\Gamma \sim 1.5 \pm 0.2$ for the young stars, but an even 
shallower distribution for the late-type (old) stars 
with $\Gamma \sim 0.2 \pm 0.1$. 
A detailed discussion concerning the different 
populations and their distribution is given in 
\cite{genzel2003stars, schoedel2007, buchholz2009, do2009stars} and 
\cite{bartko2010}.

The small value we obtain for the projected diffuse light 
exponent $\Gamma_{\mathrm{diffuse}}$
and the high degree of completeness reached around  $K_\mathrm{s}=17.5$, 
makes this data set well suited for analyzing the diffuse background 
light. 
Especially in investigating the role of much fainter stars, beyond the completeness 
limit, in the observed power-law behavior 
of the background.

\subsection{Extrapolating the KLF of the S-star cluster}
\label{subsection:extraKLF}
Motivated by the power-law behavior of the diffuse 
background emission and assuming that the drop in the 
KLF counts at magnitude~$\sim 18$ is caused only by  
the fact that we have reached the detection limit, 
we extrapolate the KLF to fainter magnitudes in order to 
investigate how these faint stars contribute to the background light.
The true shape of the luminosity function for $K_\mathrm{s}$-magnitudes 
below the completeness limit of $\sim 17.5$ has yet to be determined. Investigations 
into the IMF of the S-cluster have shown that it can be fitted with a standard 
Salpeter/Kroupa IMF of $d N/d m \propto m^{-2.3}$ and continuous star formation histories
with moderate ages  \citep[below 60~Myr,][]{bartko2010}. Here, we estimate an upper limit on 
the stellar light by assuming that the KLF exhibits the same behavior observed 
for brighter magnitudes without suffering a 
break in the slope toward the fainter end.

We use the KLF slope we obtained for the innermost central region, 
$0.18 \pm 0.07$ (Figure~\ref{klf}) and extrapolate it over five 
magnitudes bins to $K_\mathrm{s} \sim 25$. The $K_\mathrm{s}$-magnitude bins 
between 18--25 
(translating to stellar masses in the range of $\sim$~1.68 to 0.34~$M_\odot$)
correspond to the brightness of the expected 
main-sequence stars (luminosity class V) which are likely to be 
present in the central cluster. However, 
we assume that due to mass segregation effects
in the Galactic nucleus \citep{bahcall1976, alexander2005},
driven by dynamical friction \citep{chandrasekhar1943}
between stars, the heavier objects sink towards the 
center while the lighter objects move out.
Their volume density  will be significantly reduced 
and they may even be expelled from the very center.
\cite{freitag2006} show that the main-sequence 
stars begin to be expelled outward by the cusp of 
stellar-mass black holes (SBH) after a few Gyrs, 
just shorter than the presumed age of the stellar 
cluster at about 10~Gyrs.
While the reservoir of lower mass stars may be 
replenished by the most recent - possibly still 
ongoing - star formation episode about 6~million~years
 ago \citep{paumard2006}, we assume that 
stars well below 
our low mass limit of $\sim$~0.34~$M_\odot$
with $K_\mathrm{s}$-band brightnesses
around $K_\mathrm{s}=25$ are affected by depletion.

\begin{figure}[!ht]
  \centering
  \includegraphics[width=0.45\textwidth]{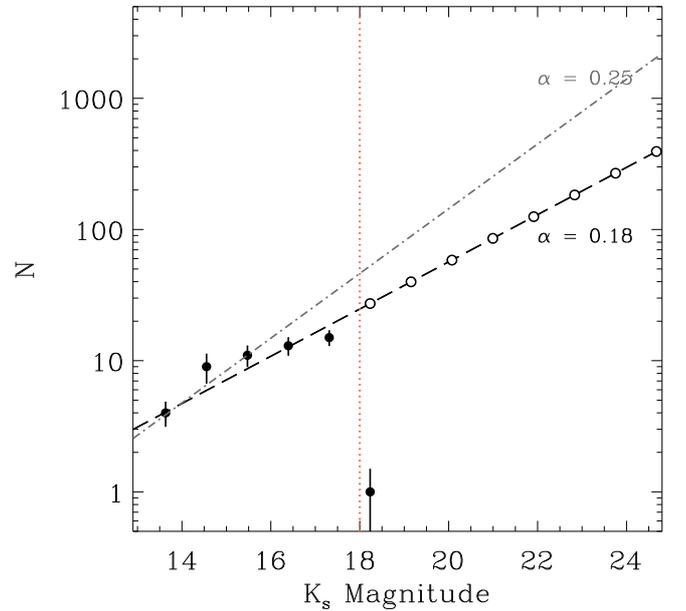}
  \caption{\small
  Extrapolation of the KLF power-law fit. The KLF slope of $\alpha= 0.18$ 
and the upper limit imposed by the uncertainty in the fit ($\alpha= 0.25$) 
are plotted as dashed and dash-dotted lines, respectively.
The black filled circles represent the data while the hollow 
circles represent new points based on the extrapolated 
KLF slope. The approximate location of the detection limit is 
indicated by the vertical dotted/red line. 
  }
  \label{klf_extra}    
\end{figure}

Figure~\ref{klf_extra} shows the KLF slope of $\alpha= 0.18$ and the upper 
limit imposed by the uncertainty in the fit ($\alpha= 0.25$) 
plotted as dashed and dash-dotted lines, respectively. 
The extrapolated $K_\mathrm{s}$-bins are shown as hollow circles. 
We adopt a Monte Carlo approach for calculating the number 
of stars $N$ from the KLF, taking into account the uncertainty in the slope. 
After $10^5$ trials we find as a result for each bin, 
the median number $N$ and median deviation $dN$.

Using the extrapolated $K_\mathrm{s}$-magnitudes, the corresponding flux 
densities are calculated using the following relation 

\begin{equation}
f_{\mathrm{new\ star}} = f_{\mathrm{S2}} \times 10^{-0.4~(K_{\mathrm{new\ star}} - K_{\mathrm{S2}})}~,
\label{eqn:fluxcalc}
\end{equation}

where $f_{\mathrm{new\ star}}$ and $K_{\mathrm{new\ star}}$ are 
the flux density and $K_\mathrm{s}$-magnitude for each new star in the extrapolation. 
The flux and magnitude for the star S2 were adopted from NS10, Table~3,
and corrected for the extinction value we use here (see \S~\ref{Observations}). 
The new values are $f_{\mathrm{S2}}= 14.73$~mJy
and $K_{\mathrm{S2}}=14.1$.
The accumulative flux density for each $K_\mathrm{s}$-bin 
$f_{\mathrm{bin}}$ is obtained via 
\begin{equation}
f_{\mathrm{bin}}= f_{\mathrm{new\ star}} \times N_{\mathrm{new\ star}}.
\end{equation}
The number of stars per bin $N_{\mathrm{new\ star}}$ is randomly picked 
from the interval between
$[N_{\mathrm{new\ star}} - dN_{\mathrm{new\ star}}]$ and 
$[N_{\mathrm{new\ star}} + dN_{\mathrm{new\ star}}]$.
In $10^5$ trials 
the accumulative flux per bin and its uncertainty 
are determined as the median and median deviation of the randomly drawn 
fluxes $f_{\mathrm{bin}}$.
We then add up all the accumulative flux densities for all the new $K_\mathrm{s}$-band bins
and obtain the integrated brightness of the extrapolated part of the S-star cluster,
\begin{equation}
F_{\mathrm {Extra\ Stars}} = \sum_{K_\mathrm{s} \simeq 18}^{ 25} f_{\mathrm{bin}} = (25.72 \pm 14.31)\quad \mathrm{mJy}.
\end{equation}
 
\begin{figure}[!ht]
 \centering
 \includegraphics[width=0.45\textwidth]{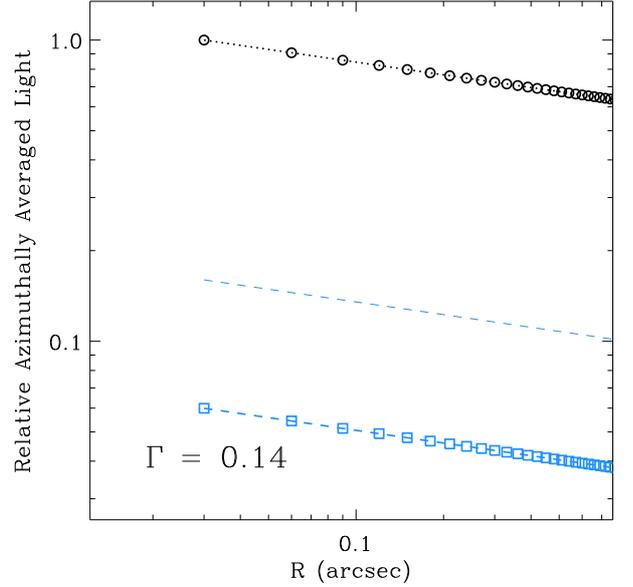}
 \caption{\small
 Relative azimuthally averaged light density, for the background 
light taken from the observations and the extra stellar light 
calculated from the extrapolation, plotted as a function of 
distance from Sgr~A*. They are represented by a dotted line with circles (black) 
and dashed lines with squares (blue), respectively.
The stellar light density is normalized to the peak light density 
of the background at the central resolution element. 
The upper limit of the extrapolated extra stellar light 
is shown as the blue dashed line with no symbols.
 }
 \label{intensity_extra}    
\end{figure}

We assume that the faint, undetectable stars 
follow the distribution of the azimuthally averaged background 
light, as shown in Figure~\ref{intensity}. 
Thus, the light from the faint stars that we introduced in the 
$0.69''$ radius region can be compared to the measured background 
light from our data for the same region. 
This is achieved by using the 
total flux density $F_{\mathrm {Extra\ Stars}}$ 
to derive the peak light density ($I_{\mathrm {Extra\ Stars}}$) 
that would be measured inside one resolution 
element of $0.033''$ radius centered on the position of Sgr~A*,
using the following relation:
\begin{eqnarray}
F_{\mathrm {Extra\ Stars}} &=& \int{f(r,\phi) r dr d\phi} \nonumber \\
                           &=& 2\pi I_{\mathrm {Extra\ Stars}} \int_{0.033^{''}}^{0.690^{''}}{r^{1-\Gamma}dr}, 
\label{eqn:peakLight}
\end{eqnarray}
with $\Gamma=\Gamma_\mathrm{diffuse} =0.14$ (see \S~\ref{subsection:DiffuseLight}).
The peak light density for the extra stars is then 
$I_{\mathrm {Extra\ Stars}}= (15.24 \pm 8.48)$~mJy~arcsec$^{-2}$.
To compare the light caused by the extra stars with the measured 
background emission, we plot 
the stellar light density caused by our new stars 
with the azimuthally averaged measured light density of the background (Figure~\ref{intensity_extra}). 
For illustration purposes we normalize the observed peak stellar light 
to the measured background value within the central 
resolution element, 
$I_{\mathrm {Background}}= (254.30 \pm 58.45)$~mJy~arcsec$^{-2}$. 
It is clear that the peak light introduced by the new faint stars, 
as calculated from the extrapolation of the $0.18 \pm 0.07$ KLF slope, 
is very small and below that of the background. 
The dotted line (black circles) represents the background light 
while the dashed line (blue squares) corresponds to the extra 
stellar light. 
The upper limit of the extrapolated extra stellar light contribution
is presented as a dashed line with no symbols.
The figure shows that the upper limit of the extrapolated 
light contribution of the S-star cluster
is lower than  $15\%$ of the measured background light.


\subsection{Observational limits on the stellar light and mass}
\label{subsection:Limits}

Our analysis shows that if there was a population of very 
faint stars, following the extrapolated $K_\mathrm{s}$-band luminosity function 
and central cluster profile obtained
for the brighter stellar population (less than $K_\mathrm{s}=18$),
the additional stellar light and mass lie well below the 
limits given by observational data. See following sections and  
Figures~\ref{inten_all} and \ref{mass_all}. 

\subsubsection{Limits on the stellar light}
\label{subsubsection:Light}
Following the previous calculations and the result displayed 
in Figure~\ref{intensity_extra}, we perform our analysis for a range 
of KLF slopes in order to test if the observed background 
light can be solely obtained by the emission of faint stars. 
The range of KLF slopes we use is based on the values and 
uncertainty estimates of the following published KLF slopes 
for the central $2''$: $0.13 \pm 0.02$ \citep[early-type stars]{buchholz2009}, 
$0.27 \pm 0.03$ \citep[late-type stars]{buchholz2009}, 
$0.21 \pm 0.02$ \citep[all stars]{buchholz2009} 
and $0.30 \pm 0.1$ (NS10), in addition to the improved newly 
fitted slope of the KLF in this work $0.18 \pm 0.07$. 

\begin{figure}[!ht]
  \centering
  \includegraphics[width=0.45\textwidth]{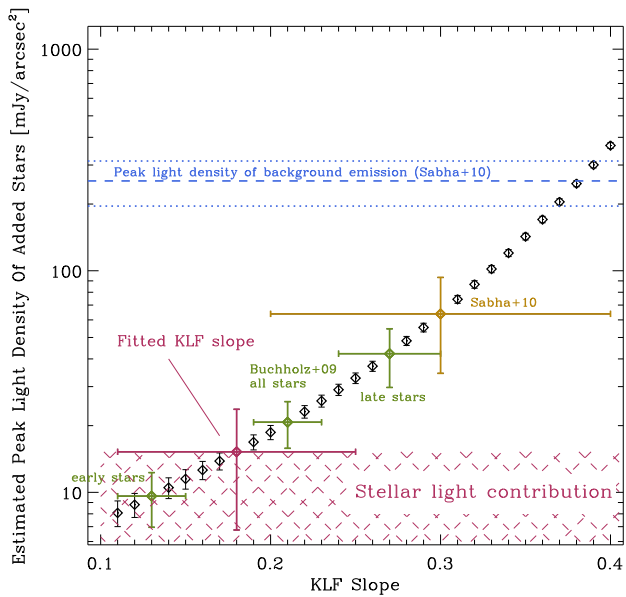}
  \caption{\small
  Estimated peak light density from stars derived from
for different KLF slopes. Slopes of \cite{buchholz2009} 
(for different stellar populations) are shown in green, 
NS10 in yellow and the KLF slope derived here, in \S~\ref{subsection:KLF}, in purple.
A limit imposed by the 
measured peak light density from the measured background 
light is plotted as a horizontal dashed line (blue). 
  }
  \label{inten_all}    
\end{figure}

We extrapolate each KLF slope to a 
$K_\mathrm{s}$-magnitude of $\simeq 25$. 
The peak light density ($I_{\mathrm {Extra\ Stars}}$) 
is calculated using 
Equation~(\ref{eqn:peakLight}). 
The peak light density of the extra 
stars is plotted for the extrapolated KLF slopes in the
range of 0.11 to 0.40 in Figure~\ref{inten_all} . 
The limit imposed by the peak light density of the
measured background light (Figure~\ref{GCmap}) is plotted as a 
horizontal dashed line (blue). 
In addition, the KLF slopes derived in this work and by NS10 and 
\cite{buchholz2009} are plotted as purple, yellow and green data points, 
respectively.  
Figure~\ref{inten_all} clearly shows that almost all of the KLF slopes
result in a peak light density below 
the observed limit, 
except for very high slopes $> 0.37$ which are not in 
agreement with the observations. 

\subsubsection{Limits on the stellar mass}
\label{subsubsection:Mass}
Using the same range of KLF slopes, we estimate the mass 
that would be introduced to the central region as a result of the
KLF extrapolation. 
We obtain the stellar mass corresponding to the extrapolated $K_\mathrm{s}$-bins by calculating their luminosity via 
\begin{equation}
L_{K_\mathrm{s}} = 10^{-0.4(M_{K_\mathrm{s}} -M_{\odot K_\mathrm{s}})} L_{\odot K_\mathrm{s}},
\end{equation}
where, $L_{K_\mathrm{s}}$ and $M_{K_\mathrm{s}}$ are the luminosity of a star and its absolute magnitude in $K_\mathrm{s}$-band, respectively.  $L_{\odot K_\mathrm{s}}$ \& $M_{\odot K_\mathrm{s}}$ are the $K_\mathrm{s}$ luminosity and absolute magnitude of the Sun. Then, the mass for each $K_\mathrm{s}$ magnitude is calculated using
\begin{equation}
m= (L_{K_\mathrm{s}})^{(1/4)}
\end{equation}
from \cite{duric2004, salaris2005}. For example, a $K_\mathrm{s}$-magnitude around 20 corresponds to 1~\solm  main-sequence stars of F0V, G0V, K5V spectral types. 

\begin{figure}[!ht]
  \centering
  \includegraphics[width=0.45\textwidth]{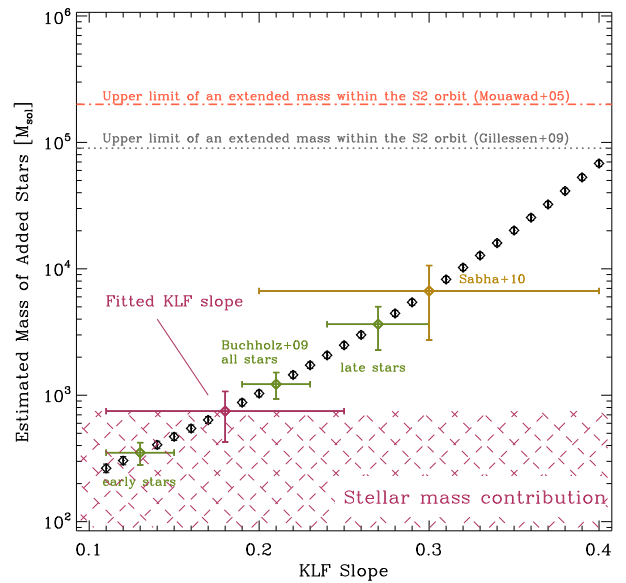}
 \caption{\small
  Estimated stellar mass from the added stars for different 
KLF slopes. Slopes of \cite{buchholz2009} (for different stellar populations)  
are shown in green, NS10 in yellow and our fitted slope in purple. 
A limit imposed by the enclosed mass within the S2 orbit is plotted 
as horizontal dotted (gray) and dash-dotted (red) lines 
from \cite{gillessen2009stars} and \cite{mouawad2005}, respectively.
  }
  \label{mass_all}    
\end{figure}

In Figure~\ref{mass_all} 
we show the estimated extra mass for all the KLF slopes 
in units of solar mass. The figure also shows, dash-dotted/red line ,
the upper limit for an extended mass enclosed by the orbit of the star S2, 
calculated by \cite{mouawad2005}, where they use non-Keplerian fitting of the 
orbit to derive the upper limits, assuming that the composition of 
the dark mass is sources with $M/L \sim2$. The dotted/gray line 
represents the tighter upper limit obtained later by \cite{gillessen2009stars}
who derive the mass using recent orbital parameters of S2.
They assume that the extended mass consists of stellar black holes 
\citep{freitag2006} with a mass of 10~\solm using estimations from  
\cite{timmes1996} and \cite{alexander2007}.  
It can be concluded from the figure that the introduced stellar mass, 
within a radius of $\sim 0.69''$, 
lies well below the upper limits imposed by the S2 orbit with a 
semi-major axis of $\sim 0.123''$ \citep{gillessen2009stars}. 
See Figure~\ref{GCmap} (right) for a comparison of the sizes of the two regions. 


\section{Dynamical probes of the distributed mass}
\label{section:DarkMass}

If the gravitational force near Sgr A* includes contributions
from bodies other than the SMBH, the orbits of test stars, including S2, 
will deviate  from Keplerian ellipses.
These deviations can be used to constrain the amount of distributed
mass near Sgr~A* \citep{mouawad2005,gillessen2009stars}.
But they can also be used to constrain the  ``granularity'' of
the perturbing potential, since the nature and magnitude of the orbital
deviations depend both on the total mass of the perturbing stars, 
and on their individual masses.

Investigations of a single scattering event were explored by \cite{gualandris2010}
using high-accuracy N-body simulations and orbital fitting techniques. 
They found that an IMBH more massive than $10^3$~$M_\odot$, with a distance 
comparable to that of the S-stars, will cause perturbations of the 
orbit of S2 that can be observed after the next 
peribothron\footnote{Peri- or apobothron is the term used for peri- or apoapsis for an elliptical orbit with a black hole present at the appropriate focus.} 
passage of S2.
Here we examine the effect many scatterers (i.e. smaller masses for the scatterers but 
shorter impact parameters) will 
have on the trajectory of the star S2 as it orbits. Around Sgr~A*, the stars and scatterers
are moving in a potential well that is dominated by the mass of the central SMBH.
In this case the encounters are of a correlated nature and hence cannot be considered as random events.

An important deviation from Keplerian motion occurs as a result of
relativistic corrections to the equations of motion, which to lowest
order predict an advance of the argument of peribothron, $\omega$,
each orbital period of
\begin{equation}\label{eq:nuGR}
\left(\Delta\omega\right)_\mathrm{GR} 
= \frac{6\pi GM_\bullet}{c^2a(1-e^2)}.
\end{equation}
Setting $a = 5.0$~mpc and $e= 0.88$ for the semi-major axis and eccentricity of S2, respectively, and assuming $M_\bullet=4.0\times 10^6 $~$M_\odot$,
\begin{equation}\label{Equation:DomegaGR}
(\Delta\omega)_\mathrm{GR} \approx 10.8^\prime.
\end{equation}
The relativistic precession is prograde, and leaves the orientation of the orbital
plane unchanged.

The argument of peribothron also experiences an advance each period
due to the spherically-symmetric component of the distributed mass.
The amplitude of this ``mass precession'' is
\begin{equation}\label{eq:nuM}
\left(\Delta\omega\right)_\mathrm{M} = -2\pi G_\mathrm{M}(e,\gamma)\sqrt{1-e^2}\left[\frac{M_\star(r<a)}{M_\bullet}\right].
\end{equation}
Here, $M_\star$ is the distributed mass within a radius $r=a$, and $G_\mathrm{M}$ is
a dimensionless factor of order unity that depends on $e$ and on the power-law index
of the density, $\rho\propto r^{-\gamma}$ \citep{Merritt2012}.
In the special case $\gamma=2$,
\begin{equation}
G_\mathrm{M} = \left(1+\sqrt{1-e^2}\right)^{-1}
\approx 0.68\ \mathrm{for\ S2}
\end{equation}
so that
\begin{equation}\label{Equation:DomegaM}
\left(\Delta\omega\right)_\mathrm{M} \approx 
-1.0^\prime\; \left[\frac{M_\star(r<a)}{10^3 \mathrm{~M}_\odot}\right].
\end{equation}
Mass precession is retrograde, i.e., opposite in sense to the relativistic precession.

Since the contribution of relativity to the peribothron advance is determined 
uniquely by $a$ and $e$, which are known,
a measured $\Delta\omega$ can be used to constrain the 
mass enclosed within S2's orbit, by subtracting
$(\Delta\omega)_\mathrm{GR}$ and comparing the result with Equation~(\ref{Equation:DomegaM}).
So far, this technique has yielded only upper limits on $M_\star$ of 
$\sim 10^{-2}M_\bullet$ \citep{gillessen2009stars}.

The granularity of the distributed mass makes itself felt via the phenomenon
of ``resonant relaxation'' (RR) \citep{RT96,HA2006}.
On the time scales of interest here,
orbits near Sgr~A* remain nearly fixed in their orientations, 
and the perturbing effect of each field star on the motion of a test star
(e.g. S2) can be approximated 
as a torque that is fixed in time, and proportional to $m$, the mass of the field star.
The net effect of the torques from $N$ field stars is to change the angular momentum, $\mathbf{L}$, of S2's orbit according to
\begin{equation}\label{Equation:dLRR}
\frac{|\Delta\mathbf L|}{L_c} \approx K\sqrt{N}\frac{m}{M_\bullet}\frac{\Delta t}{P}
\end{equation}
where $L_c=\sqrt{GM_\bullet a}$ is the angular momentum of a circular 
orbit having the same semi-major axis as that of the test star.
(Equation~\ref{Equation:dLRR} describes ``coherent resonant relaxation'';
on time scales much longer than orbital periods, 
``incoherent'' resonant relaxation causes changes that increase as $\sim \sqrt{\Delta t}$.)
The normalizing factor $K$ is difficult to compute from first principles but
should be of order unity \citep{Eilon2009}.
Changes in $\mathbf{L}$ imply changes in both the eccentricity, $e$, of S2's
orbit, as well as changes in its orbital plane.
The latter can be described in a coordinate-independent way via the angle
$\Delta\theta$, where
\begin{equation}
\cos (\Delta\theta) = \frac{\mathbf{L}_1\cdot\mathbf{L}_2}{L_1L_2}
\end{equation}
and $\{\mathbf{L}_1, \mathbf{L}_2\}$ are the values of $\mathbf{L}$ at two
times separated by $\Delta t$.
If we set $\Delta t$ equal to the orbital period of the test star,
the changes in its orbital elements due
to RR are expected to be 
\begin{eqnarray}
|\Delta e|_\mathrm{RR} &\approx& K_e \sqrt{N}\frac{m}{M_\bullet},
\label{Equation:RRa} \\ 
(\Delta \theta)_\mathrm{RR} &\approx& 2\pi K_t \sqrt{N}\frac{m}{M_\bullet}, 
\label{Equation:RRb}
\end{eqnarray}
where $N$ is the number of stars having $a$-values
similar to, or less than, that of the test star and $\{K_e, K_t\}$ are 
constants which may depend on the 
properties of the field-star orbits.

Because the changes in S2's orbit due to RR scale differently with $m$ and
$N$ than the changes due to the smoothly-distributed mass,
both the number and mass of the perturbing objects within S2's orbit 
can in principle be independently constrained.
For instance, one could determine $M_\star=mN$ from Equations~(\ref{eq:nuGR})
and (\ref{Equation:DomegaM}) and a measured $\Delta\omega$, then
 compute $m\sqrt{N}$ by measuring changes in $e$ or $\theta$ and comparing
with Equations (\ref{Equation:RRa}) or (\ref{Equation:RRb}).

We tested the feasibility of this idea using numerical integrations.
The models and methods were
similar to those described in \cite{merritt2010RR}.
The $N$ field stars were selected from a density profile $n(r) \propto r^{-2}$,
with semi-major axes extending to $a_\mathrm{max}=8$~mpc.
Initial conditions assumed isotropy in the velocity distribution.
Two values for the field star masses were considered:
$m=10$~\solm and $m=50$~$M_\odot$. One of the $N$-body particles was assigned the observed mass and
orbital elements of S2;
this particle was begun at apobothron, and the integrations extended for
one complete period of S2's orbit.
Each of the $N$ field-star orbits were integrated as well, and the
integrator included the mutual forces between stars, as well as post-Newtonian
corrections to the equations of motion.
The quantities $\Delta\omega$, $\Delta e$ etc. for the S2 particle were computed by
applying standard formulae to $(\mathbf{r}, \mathbf{v})$ at the start
and end of each integration.
100 random realizations of each initial model were integrated, allowing both
the mean values of the changes, and their variance, to be computed.
\begin{figure}[!t]
  \centering
  \includegraphics[width=0.45\textwidth]{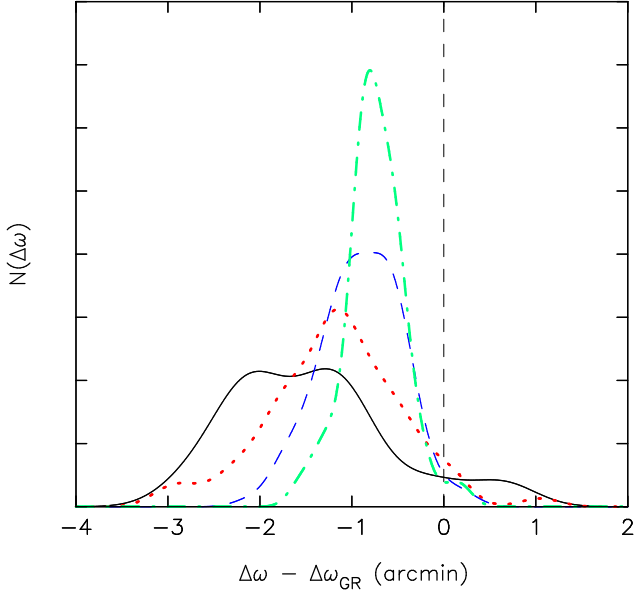}
  \caption{\small
Histograms of the predicted change in S2's argument of peribothron, $\omega$,
over the course one orbital period ($\sim16$~yr).
The shift due to relativity, $(\Delta\omega)_\mathrm{GR}\approx 11^\prime$, 
has been subtracted from the total; what remains is due to 
Newtonian perturbations from the field stars.
Each histogram was constructed from integrations of 100 random realizations of the
same initial model, with field-star mass $m=10$~$M_\odot$, and four
different values of the total number: $N=200$ (solid/black);
$N=100$ (dotted/red); $N=50$ (dashed/blue); and $N=25$ 
(dot-dashed/green).
The average value of the peribothron shift increases with increasing
$Nm$, as predicted by Equation~(\ref{Equation:DomegaM}).
The reasons for the spread in $\Delta\omega$ values are discussed in the text.
}
  \label{S2histogram}    
\end{figure}
\begin{figure}[!ht]
  \centering
  \includegraphics[width=0.45\textwidth]{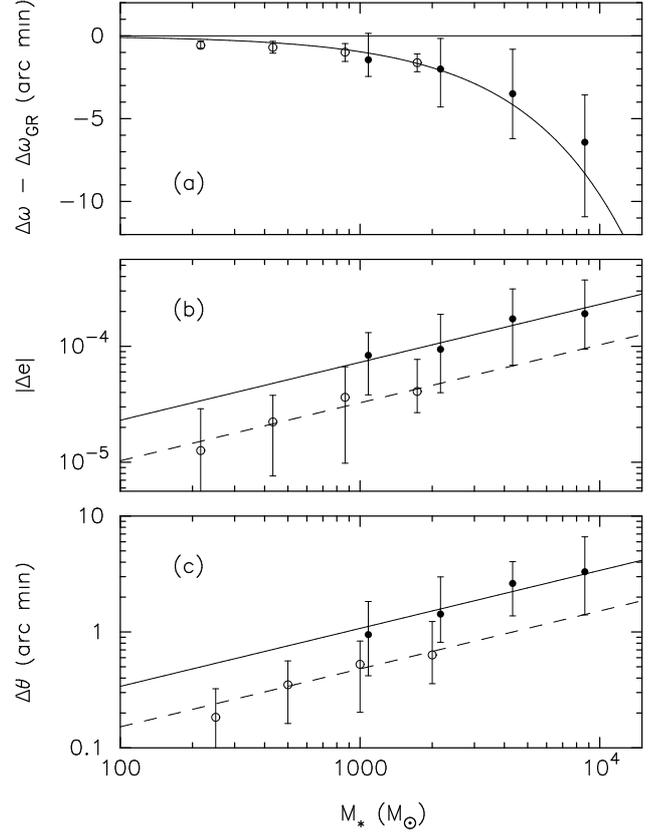}
  \caption{\small
Average values of the changes in $\omega$, $e$ and $\theta$ for S2
over one orbital period ($\sim16$~yr) in the $N$-body integrations.
Filled circles are from integrations with $m=50$~\solm and open circles
are for $m=10$~\solm; the number of field stars was $N=\{25,50,100,200\}$
for both values of $m$.
The abscissa is the distributed mass within S2's apobothron, at $r\approx 9.4$~mpc.
In each frame, the points are median values from the 100 $N$-body integrations,
and the error bars extend from the $20$th to the $80$th percentile of the distribution.
\textbf{a}) Changes in the argument of peribothron. The contribution from relativity,
Equation~(\ref{eq:nuGR}), has been subtracted.
The solid line is Equation~(\ref{Equation:DomegaM}).
\textbf{b}) Changes in the eccentricity.
Solid and dashed lines are Equation~(\ref{Equation:RRa}), with $m=50$~\solm and
$m=10$~\solm respectively and with $K_e=1.4$.
\textbf{c}) The angle between initial and final values of $\mathbf{L}$ for S2.
Solid and dashed lines are Equations~(\ref{Equation:RRb})
with $K_t=1.0$.
  }
  \label{S2changes}    
\end{figure}

Figures~\ref{S2histogram} and \ref{S2changes}a show changes in $\omega$ for S2.
The median change is well predicted by Equation~(\ref{Equation:DomegaM}).
However there is a substantial variance.
We identify at least two sources for this variance.
(1) The number of stars inside S2's orbit differs from model to model by
$\sim\sqrt{N}$, resulting in corresponding changes to the enclosed mass,
and hence to the precession rate as given by Equation~(\ref{Equation:DomegaM}).
(2) When $N$ is finite, the same torques that drive resonant relaxation also imply
a change in the field star's rate of peribothron advance as compared with Equation~(\ref{Equation:DomegaM}), which assumes no tangential forces. 
While the dispersion scales roughly as $\sqrt N$, as evident in Figure~\ref{S2histogram}, 
the fractional change in $\Delta\omega$ due to this effect scales as $\sim 1/\sqrt{N}$ \citep{merritt2010RR}.
Additional variance might arise from close encounters between field stars
and S2, and from the fact that the mass within S2's orbit is changing over
the course of the integration due to the orbital motion of  each field star.

Whereas the (average) value of $\Delta\omega$ depends only on the mass
within S2's orbit, the changes in $e$ and $\theta$ depend also on $m$, as shown
in Figures~\ref{S2changes}b and c.
The lines in those figures are Equations~(\ref{Equation:RRa}) and
(\ref{Equation:RRb}), with
\begin{equation}
K_e=1.4, \ \ \ \ 
K_t = 1.0.\nonumber
\end{equation}
(We have defined $N$ in Equations~(\ref{Equation:RRa}) and (\ref{Equation:RRb})
as the number of field stars 
inside a radius of $9.4$~mpc, the apobothron of S2.)
For a given value of the enclosed mass, 
$M_\star=Nm$, Figure~\ref{S2changes}
shows that the changes in $e$ and $\theta$ indeed scale as $\sim1/\sqrt{N}$
or as $\sim\sqrt{m}$,
as predicted by Equations~(\ref{Equation:RRa}) and (\ref{Equation:RRb}).

We can use these results to estimate the changes in $\omega$, $e$
and $\theta$ expected for S2, based on theoretical models of the distribution of
stars and stellar remnants at the GC.
In dynamically evolved models \citep{freitag2006,hopman2006}, 
the total distributed mass within S2's apobothron, $r\approx 10$~mpc, is
predicted to be $\sim$ a few times $10^3$~$M_\odot$.
About half of this mass is in the form of main-sequence stars
and half in stellar-mass black holes, with a total number 
$N\approx 10^3$.
When there are two mass groups, expressions like Equations~(\ref{Equation:RRa}) and (\ref{Equation:RRb}) generalize to
\begin{eqnarray}
|\Delta e|_\mathrm{RR} &=& K_e \left[\frac{m_1\sqrt{N_1} + m_2\sqrt{N_2}}{M_\bullet}\right]\\
(\Delta \theta)_\mathrm{RR} &=& 2\pi K_t \left[\frac{m_1\sqrt{N_1} + m_2\sqrt{N_2}}{M_\bullet}\right]
\end{eqnarray}
assuming
\begin{equation}
\nonumber m_1 = 1 \mathrm{~M}_\odot, \quad m_2 = 10 \mathrm{~M}_\odot, \quad N_1 = 10^3 , \quad N_2 = 150 
\end{equation}
\citep{hopman2006}
we find
\begin{eqnarray}\label{Equation:Depredicted10^3}
|\Delta e|_\mathrm{RR} &\approx& 5.4 \times 10^{-5}, \\
(\Delta\theta)_\mathrm{RR} &\approx& 0^\prime.8, \label{Equation:Dtpredicted10^3}\\ 
(\Delta\omega)_\mathrm{M} &\approx& -2.5'.\label{Equation:Dopredicted10^3}
\end{eqnarray}
For obtaining the dispersion in the value of Equation~(\ref{Equation:Dopredicted10^3}), we scaled the dispersion given in Figure~\ref{S2changes}a for the single population case, $N=50, m=50$~\solm of the same total extended mass, to the two populations case we are investigating here. The dispersion obtained from the simulations is  $\sim 4'$. We scale it using the relation $\Delta \omega /\sqrt{N}$ in order to account for the SBH and MS populations, independently. The dispersion for the new configuration then becomes $\sim 1.43' $, lower than the single population case. This is attributed to the fact that the number of main-sequence stars is much larger than the stellar-mass black holes, hence they lower the dispersion in the total Newtonian peribothron shift $(\Delta\omega)_\mathrm{M}$.   

Considering a higher value for the enclosed mass $M_\star=10^4$~\solm while keeping the same mass scales and abundance ratios
of the scattering objects,
\begin{equation}
\nonumber m_1 = 1 \mathrm{~M}_\odot, \quad m_2 = 10 \mathrm{~M}_\odot, \quad N_1 = 4000 , N_2 = 600 
\end{equation} 
one gets changes of 
\begin{eqnarray}\label{Equation:Depredicted10^4}
|\Delta e|_\mathrm{RR}        &\approx&     1.1 \times 10^{-4}, \\
(\Delta\theta)_\mathrm{RR}    &\approx&     1.7', \label{Equation:Dtpredicted10^4}\\
(\Delta\omega)_\mathrm{M} &\approx& -10'.\label{Equation:Dopredicted10^4}
\end{eqnarray}
The dispersion in Equation~(\ref{Equation:Dopredicted10^4}) can be compared, as we did before, to the case considered in the simulations ($N=200, m=50$~$M_\odot$)  by scaling the $\sim 8'$ dispersion (Figure~\ref{S2changes}a) to become $\sim 2.86'$ for the two mass population.

Repeating the same analysis as before to the $M_\star=10^5$~\solm gives the following numbers for the stellar black holes and low-mass stars
\begin{equation}
\nonumber m_1 = 1 \mathrm{~M}_\odot, \quad m_2 = 10 \mathrm{~M}_\odot,  N_1 = 40000 , N_2 = 6000 
\end{equation} 
that result in 
\begin{eqnarray}\label{Equation:Depredicted10^5}
|\Delta e|_\mathrm{RR}        &\approx&     3.4 \times 10^{-4}, \\
(\Delta\theta)_\mathrm{RR}    &\approx&     5.2', \label{Equation:Dtpredicted10^5}\\
(\Delta\omega)_\mathrm{M} &\approx& -100'. \label{Equation:Dopredicted10^5}
\end{eqnarray}
Similar to the above cases, the dispersion in Equation~(\ref{Equation:Dopredicted10^5}) can be compared to the single mass
case by scaling the  $\sim 25.3'$ dispersion to become $\sim 9.1'$ for the two mass population. The $\sim 25.3'$ value is
obtained by scaling with $\Delta \omega /\sqrt{N}$ from the value shown in Figure~\ref{S2changes}a for the $10^4$~\solm extended mass.

We would like to stress that making a definite prediction about the $N$-dependence of the variance
is beyond the scope of the current paper.
However, we have noted that in both cases considered in Figure~\ref{S2changes} the relative variance 
is of the order 
of unity or larger i.e. the dispersion is of the
order of the Newtonian peribothron shift. 

The positional uncertainty is currently of the order of 1~mas. For the highly eccentric orbit of S2 this implies that the accuracy with which the peribothron shift can be detected is of 
the order of $24'$. As can be seen for the case of $M_\star = 10^5$~\solm, the shifts are at the limit of the current instrumental capabilities if the total enclosed mass was entirely composed of massive perturbers. The shifts given in Equations~(\ref{Equation:Dopredicted10^3}) and (\ref{Equation:Dopredicted10^4}) can be measured if the accuracy is improved by at least one order of magnitude using larger telescopes or interferometric methods in the NIR. However, considering the variances in the calculated shifts one would need to observe more than one stellar orbit in order to infer information on the population giving rise to the Newtonian peribothron shift.
By comparison, the current uncertainty in S2's eccentricity is $\sim 0.003$,
and  uncertainties in the Delaunay angles $i$ and $\Omega$ describing its orbital plane are
$\sim 50^\prime$ \citep{gillessen2009stars}.
In both cases, an improvement of a factor $\sim 50$ would be required
in order to detect the changes given in $e$ and $\theta$.

Dynamically-relaxed models of the GC have been criticized
on the grounds that they predict a steeply-rising density of old stars inside
$\sim 1$~pc, while the observations show a parsec-scale core
\citep{buchholz2009,do2009stars,bartko2010}.
Dynamically {\it un}relaxed models imply a much 
lower density near Sgr A* and an uncertain fraction of stellar-mass
black holes \citep{merritt2010,antonini2012}.
The number of perturbers is so small in these models that their effect on the
orbital elements of S2 would be undetectable for the foreseeable future,
barring a lucky close encounter with S2.

In addition to the small amplitude of the perturbations,
the potential difficulty in constraining $N$ and $m$
comes from the nonzero variance of the predicted changes (Figure~\ref{S2changes}).
The variance in $\Delta\omega$ scales as $\sim \Delta\omega/\sqrt{N}$ and
would be small in the dynamically-relaxed models with $N\approx 10^3$. 
Another source of uncertainty comes from the dependence of 
the amplitude of $\Delta \omega$ on $\gamma$ (Equation~\ref{eq:nuM}), which is unknown.
We do not have a good model for predicting the variances in $|\Delta e|$ 
and $\Delta\theta$, but Figure~\ref{S2changes} suggests that the fractional
variance in these quantities
is not a strong function of $N$ or $m$, and that it is large enough to 
essentially obscure changes due to a factor $\sim 5$ change in $m$
at fixed $M_\star$.
On the other hand, considerably more information might be available than
just $\Delta e$ and $\Delta\theta$ for one star; for instance, 
the full time-dependence of $(\mathbf{r}, \mathbf{v}$) for a number of stars.
We leave a detailed investigation of how well such information
could constrain the perturber $m$ and $N$ to a future work.


\subsection{Fighting the limits on the power of stellar orbits}
\label{subsection:OrbitLimits}

The results from the previous sub-sections clearly show that
deriving the net-displacement for an ideal elliptical orbit
for a single star will not be sufficient to put firm limits 
on both the total amount of 
extended mass and on the nature of the corresponding
population.
However, the situation may be improved if one studies 
the statistics of the time and position dependent deviations 
along a single star's orbit or instead uses the orbits of several stars.

\subsubsection{Improving the single orbit case}
\label{subsubsection:SingleOrbit}

The actual uncertainty in projected right ascension or declination, 
$\sigma_{\mathrm{position}}^2$,
can be thought of as a combination of several contributions.
Here $\sigma_{\mathrm{apparent}}^2$ is the apparent positional 
variation due to the photo-center variations of the star while
it is moving across the sea of fore- and background sources.
The scattering process results in a variation of positions 
described by $\sigma_{\mathrm{scattering}}^2$. Finally, systematic uncertainties
due to establishing and applying an astrometric reference frame
give a contribution of $\sigma_{\mathrm{systematic}}^2$.

The value of $\sigma_{\mathrm{position}}^2$ can be measured in comparison to the
orbital fit.
The value of $\sigma_{\mathrm{apparent}}^2$ can be obtained experimentally 
by placing an artificial star into the imaging frames 
at positions along the idealized orbit. A reliable estimate of 
$\sigma_{\mathrm{apparent}}^2$ is achieved by comparing the known positions 
at which the star has been placed and the positions measured in the image frames.
As for the case of the systematic variations, they can be estimated by investigating 
sources that are significantly brighter or slower than the S-stars.
Finally, the value that describes the scattering process, and therefore gives
information on the masses of the scattering sources, can be obtained via
\begin{equation}
\sigma_{\mathrm{scattering}}^2=
\sigma_{\mathrm{position}}^2-
\sigma_{\mathrm{apparent}}^2-
\sigma_{\mathrm{systematic}}^2.
\end{equation}

Alternatively, $\sigma_{\mathrm{scattering}}^2$ could be measured directly by near-infrared
interferometry with long baselines. Measuring the position of S2 
interferometrically as
a function of time with respect to bright reference objects could allow
observing the effects of single scattering events. Here the assumption 
is that they happen infrequently enough such that one can build up 
sufficient signal to noise on the $\sigma_{\mathrm{scattering}}$ measurement
provided that the uncertainties in the interferometric
measuring process are sufficiently well known.


\subsubsection{Improving by using several stars}
\label{subsubsection:SeveralStars}

If scattering events contribute significantly to the uncertainties in 
the determination of the orbits,
a number of stars may help to derive the physical
properties of the medium through which the stars are moving.
While the influence of the extended mass imposes a systematic 
variation of the orbits through the Newtonian peribothron shift,
the variations due to scattering events will be random. This implies that
for individual stars the effects may partially compensate or amplify
each other. 
Averaging the results of $N$ stars, that will then essentially sample
the shape of the distributions shown in Figure~\ref{S2histogram},    
may therefore result in an improvement proportional to $N^{-1/2}$ 
in the determination of the extended mass.


\section{Simulating the distribution of fainter stars}
\label{section:SimulateStars}

In NS10 we detected three stars that were either previously not
identified at all (NS1 \& NS2 stars, Figure~1 in NS10) or 
only allowed an unsatisfactory identification with previously
known members of the cluster \citep[S62, as pointed out in][]{dodds-eden2011}.
In addition we have the case of the star S3 which was identified in
the $K_\mathrm{s}$-band in 
the early epochs  1992 \citep{eckart1996}, 1995 \citep{ghez1998} 
and lost after about 3 years in 1996/7 \citep{ghez1998}, 
1998 \citep{genzel2000}.
We investigate this phenomenon in our modeling by extrapolating the KLF 
in the inner 1--2~arcsec region, surrounding Sgr~A*, to stars fainter
than the faintest source ($K_\mathrm{s}=17.31$) we detected in our 
30 August and 23 September 2004 dataset, in which Sgr~A* shows very
low activity (NS10).
In this section we describe the method we use to simulate the distribution 
of these faint stars, and the possible false detections 
that can be caused by the combined light of many stars 
appearing in projection to be very close to each 
other, such that they cannot be individually resolved with 8--10~m class telescopes. 

The calculations were done by taking all the extra (extrapolated) 
faint stars in the $K_\mathrm{s}$-magnitude interval of 18 to 25.
The stars were then distributed in a $23 \times 23$ grid that corresponds to 529 cells. 
Each cell has the dimensions of $0.06'' \times 0.06''$, i.e. about 
one angular resolution 
element in $K_\mathrm{s}$-band, this grid, therefore, simulates observations of 
the inner $1.38'' \times 1.38''$ projected region surrounding Sgr~A*. 
We distributed the faint stars in the grid such that their radial 
profile centered on Sgr~A* reproduces that of the stellar number density 
counts of the inner region of the central stellar cluster with a 
power-law index of $\Gamma = 0.30 \pm 0.05$ from \cite{schoedel2007}. 
This way each cell has a specific number 
of stars that can be inserted into it, with the maximum number of 
stars being located in the central cell, i.e. the peak of the 
radial profile. Our algorithm  fills each cell with its 
specified number of stars by choosing them randomly from a pool 
of stars created from the extrapolated KLF. The pool is created such that 
for each $K_\mathrm{s}$-magnitude bin above $\sim 18$, a number of stars $N$ 
get their $K_\mathrm{s}$-band magnitudes according to the KLF.
From this pool of stars we then randomly pick objects to fill the cells 
of the grid such that they obey the power-law radial number density 
profile. 
Then, the fluxes of the stars in each individual cell are added up
and compared to the value of 
0.76~mJy
which is the flux density of
the faintest stellar source in our S-star cluster data, i.e. $K_\mathrm{s}=17.31$ (NS10).
We ran the simulation $10^4$~times 
in order to get reliable statistical estimates for the brightnesses 
in each resolution cell. Hence we can estimate how likely it is to find
strong apparent clusterings along the line of sight that are brighter than 
the faintest star we identified in the S-cluster (flux larger than 
0.76~mJy).
\begin{table}[!ht]
\centering
\caption{Probabilities of detecting a false star (brighter than $K_\mathrm{s}=17.31$) in a $1.38''\times 1.38''$ region.
\label{montecarlo}}
{\begin{small}
\newcolumntype{M}[1]{>{\raggedright}m{#1}}
\begin{tabular}{M{2.5cm}ccccc}
\hline
\hline
$K_\mathrm{s}$-band       & \multicolumn{3}{c}{Power-law index}   \\
magnitude    &              &              &             \\
cutoff       &  0.19        &  0.30        &  0.35       \\
\hline
\hline

\multicolumn{4}{c}{KLF slope $=0.11$}\\ 
	     &              &              &             \\
20.99        & $0.0000$     & $0.1471$     & $0.1500$    \\
             &              &              &             \\
24.67        & $0.0285$     & $0.0292$     & $0.0224$    \\
\hline
\multicolumn{4}{c}{KLF slope $=0.18$}\\
	     &              &              &             \\
20.99        & $0.0848$     & $0.1286$     & $0.3016$    \\
             &              &              &             \\
24.67        & $0.2058$     & $0.2426$     & $0.2927$    \\
\hline
\multicolumn{4}{c}{KLF slope $=0.25$}\\
	     &              &              &             \\
20.99        & $0.7776$     & $0.7442$     & $0.9085$    \\
             &              &              &             \\
24.67        & $0.9462$     & $0.9725$     & $0.9802$    \\
\hline

\end{tabular}
\end{small}}
\end{table}

Taking into account the uncertainties of the quantities that describe the
central S-star cluster we have repeated the simulation for a 
combination of three KLF slopes 
($0.11$, $0.18$ and $0.25$), three radial profile power-law indices 
($\Gamma =0.19$, $0.30$ and $0.35$) 
and two $K_\mathrm{s}$-magnitude cutoffs for the extrapolation,
$21$ and $25$ (corresponding to
0.0258 and 0.0009~mJy, respectively). 
Here the brighter cutoff is very close to the brightness of the 
faintest stars that have been detected.
The choice for the KLF slope satisfies 
the range of the power-law fit $\Gamma=0.18 \pm 0.07$. The power-law 
indices were taken from Table~5 of \cite{schoedel2007} for the 
cusp radial profiles. 

\begin{figure*}
  \centering
  \includegraphics[width=\textwidth,angle=00]{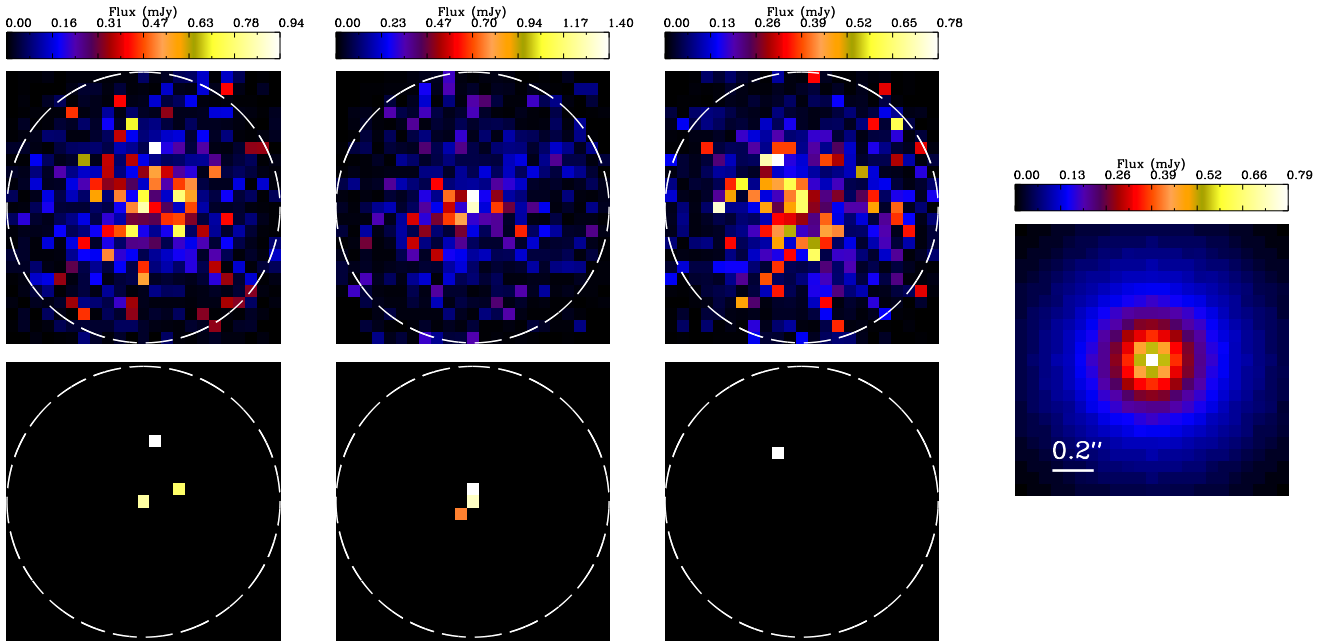}
  \caption{\small
  \textit{Upper panel}: Three different snapshots of the simulation for the 0.25
KLF slope,
  power-law index $\Gamma=0.3$ and 24.67 $K_\mathrm{s}$-magnitude cutoff. \textit{Lower
panel}: The same as upper
  but with only the detectable blend stars visible.
  \textit{Right}: Average of all the $10^4$ simulation snapshots for the same
setup. }
  \label{simulatedImage}   
\end{figure*}

The results of the simulations are summarized in Table~\ref{montecarlo}. 
Three different realizations of a cluster simulation as well as the 
average of $10^4$ simulations are shown in  Figure~\ref{simulatedImage}.  
We find that for the measured KLF slope of $0.18$, a measured power-law index
of $\Gamma=0.3$ and a faint $K_\mathrm{s}$-magnitude cutoff we obtain a false star in about a quarter of 
all simulations.
For steeper KLF and power-law slopes $\Gamma$ we get this result in more than
70\% of all cases independent of the cutoff magnitude.

\begin{table}[!ht]
\centering
\caption{Probabilities of detecting a false star (brighter than $K_\mathrm{s}=17.31$) at the position of Sgr~A*. \label{montecarlocentral}}
{\begin{small}
\newcolumntype{M}[1]{>{\raggedright}m{#1}}
\begin{tabular}{M{2.5cm}ccccc}
\hline
\hline
$K_\mathrm{s}$-band       & \multicolumn{3}{c}{Power-law index}    \\
magnitude    &              &              &              \\
cutoff       &  0.19        &  0.30        &  0.35        \\
\hline
\hline

\multicolumn{4}{c}{KLF slope $=0.11$}\\
	     &              &              &              \\
20.99        & $0.0000(2)$  & $0.0752(4)$  & $0.0757(4)$  \\
             &              &              &              \\
24.67        & $0.0094(6)$  & $0.0103(6)$  & $0.0099(6)$  \\
\hline
\multicolumn{4}{c}{KLF slope $=0.18$}\\
	     &              &              &              \\
20.99        & $0.0423(4)$  & $0.0438(4)$  & $0.1181(5)$  \\
             &              &              &              \\
24.67        & $0.0345(15)$ & $0.0591(18)$ & $0.0821(20)$ \\
\hline
\multicolumn{4}{c}{KLF slope $=0.25$}\\
	     &              &              &              \\
20.99        & $0.3120(8)$  & $0.3149(8)$  & $0.5448(10)$ \\
             &              &              &              \\
24.67        & $0.3223(59)$ & $0.4756(70)$ & $0.5291(74)$ \\
\hline

\end{tabular}
\\
\tablefoot{ The number of stars contributing to the detected flux of the false star is given in parentheses for each considered case.}
\end{small}}
\end{table}
In Table~\ref{montecarlocentral} we show the same statistics as in 
Table~\ref{montecarlo} but for the central cell in 
the grid, at the projected position of Sgr~A*. 
Also given, in parentheses, is the number of stars in the central cell that gives rise 
to the detection of a false star  at a distance of less than one angular resolution
 element away from the line of sight to Sgr~A*. 
We find that for a 
KLF slope of $0.25$ we get a false star in 30\% to 50\%
of all simulations, independent of the power-law index $\Gamma$
and the cutoff magnitude.
This is consistent with the offsets found in different observational epochs 
of Sgr~A* light curves 
\citep{witzel2012, dodds-eden2011}.
In this case the blend consists of 8 to 74 stars below
the unresolved background in the S-star cluster region.
For flatter KLF slopes (i.e. 0.11 and 0.18) 
we find that a blend star only occurs 
in less than about 10\%
of all cases, which appears to be well below the upper limit found 
from observations.
For a KLF slope of $\alpha=0.25$ and a number density power-law 
index of $\Gamma$ around 0.3 the total number of stars   
in the simulated S-star cluster is a few 1000.
This is consistent with the number of main-sequence stars assumed by 
\cite{freitag2006}.

\section{Summary and conclusion}
\label{section:Summary}

By determining the KLF of the S-star 
cluster members from infrared imaging, 
using  the distribution of the diffuse background light and the stellar number density counts, 
we have been able to shed some light on the 
amount and nature of the stellar 
and dark mass associated with the cluster of high velocity S-stars 
in the immediate vicinity of Sgr~A*.

The amount of light from
the fainter S-cluster members is below the amount of residual light 
after removing the bright 
cluster members.
One implication 
could be that both the diffuse light
and dark mass are overestimated. 
However, while NS10 estimate that only a
maximum of one third of the diffuse light could
be due to residuals from the PSF subtraction, 
we find that faint stars at or beyond
the completeness limit reached in the KLF can account 
only for about 15\% of the background light. 
Additional light may also originate from accretion
processes onto a large number of 10~\solm black holes that
may reside in the central region, covered by the S-stars.
We find that the stellar mass derived from the 
KLF extrapolation is much smaller than the amount of mass 
that may be present considering the uncertainties in the orbital 
motion of the star S2. 
Higher angular resolution and sensitivity are needed
to resolve the background light and analyze its origin.

By investigating the effects of orbital torques due to resonant relaxation, we find that if a
significant population of 10~\solm black holes is present, 
with enclosed masses between $10^3$~\solm and $10^5$~\solm 
\citep[see e.g.][]{freitag2006}, then 
for trajectories of S2-like stars, contributions from
scattering will be important compared to the 
relativistic or Newtonian peribothron shifts.
This clearly shows that observing a
single stellar orbit will not be sufficient to put firm limits on the total amount of 
extended mass and on the importance of relativistic 
peribothron shift.
In this case only the observation of a larger number of stars
will allow to sample the statistics of the effect,
i.e. the distributions in Figure~\ref{S2histogram}.
However, if the distribution of 10~\solm black holes is
cuspy then this may become even more difficult (and 
close encounters should be frequent in this region).

In general, the inclusion of star-star perturbations allows us to probe the distribution and composition 
of mass very close to the SMBH simultaneously, if
the astrometric accuracy can be improved by an order of magnitude 
by using either larger telescopes or interferometers in the NIR.

With measurements and extrapolations of the S-star cluster
KLF slope, and number density counts with assumptions on the
KLF cutoff magnitude, we can show that the contamination for the 
members of the cluster, and especially at the position of Sgr~A*,  
by blend stars is fully consistent with measurements.
We show that for 8--10~m class telescopes the presence and
proper motion of faint stars close to the confusion limit 
in the region of the S-star cluster is highly contaminated by blend stars.
Due to the 2-dimensional velocity dispersion of the stars within 
the S-star cluster of about 600~km/s the blend stars will last for 
about 3--4 years before they fade and dissolve. 
Close to the center, we find the probability of detecting blend stars at
any time is about 30--50\%. 
At the central position the change from the appearance of a blend 
star to the appearance of another may also give the
illusion of high proper motions for 8--10~m class telescopes.
Such a prime example would be S3, detected close to the position of Sgr~A*, 
which had both a limited lifetime 
and high proper motion 
\citep{eckart1996, ghez1998, genzel2000}.
Blending of sources along the line of sight 
may also severely contaminate the proper motion 
measurements of individual stars close to the confusion limit.
Only with the help of proper motion measurements 
over time significantly longer than 3 years one will be able to
derive reliable orbital parameters for a single star.
Also, spectroscopy may help to resolve blend stars, however, the objects are
faint and spectroscopy will be difficult.

These findings clearly demonstrate the necessity of higher angular 
resolution, astrometric accuracy and point source sensitivity 
for future investigations of the S-star cluster. 
They would also greatly improve the derivation 
of the amount and the compactness
of the central mass as well as the determination of relativistic 
effects in the vicinity of Sagittarius~A*. 

\begin{acknowledgements}
We thank the anonymous referee for the helpful comments.
N.~Sabha is member of the Bonn Cologne Graduate School (BCGS) for
Physics and Astronomy supported by the Deutsche Forschungsgemeinschaft.
D.~Merritt acknowledges support from the National Science Foundation under
grants no. AST 08-07910, 08-21141 and by the National Aeronautics and
Space Administration under grant no. NNX-07AH15G.
We thank Tal Alexander for useful discussions.
M.~Garc\'{\i}a-Mar\'{\i}n is supported by the German 
federal department for education and research (BMBF) under 
the project number 50OS1101. M.~Valencia-S. and B.~Shahzamanian  are members of the 
International Max-Planck Research School (IMPRS) 
for Astronomy and Astrophysics at the Universities of Bonn and Cologne
supported by the Max Planck Society.
Part of this work was supported by the German Deutsche 
Forschungsgemeinschaft, DFG, via grant SFB 956 and fruitful discussions 
with members of the European Union funded COST Action MP0905: 
Black Holes in a violent Universe and PECS project No. 98040.

\end{acknowledgements}

\bibliographystyle{aa} 
\bibliography{mybibNod} 

\begin{thebibliography}{67}
\expandafter\ifx\csname natexlab\endcsname\relax\def\natexlab#1{#1}\fi

\bibitem[{{Alexander}(2005)}]{alexander2005}
{Alexander}, T. 2005, \physrep, 419, 65

\bibitem[{{Alexander}(2007)}]{alexander2007}
{Alexander}, T. 2007, arXiv e-prints:0708.0688

\bibitem[{{Alexander} \& {Hopman}(2009)}]{alexander2009}
{Alexander}, T. \& {Hopman}, C. 2009, \apj, 697, 1861

\bibitem[{{Antonini} {et~al.}(2012){Antonini}, {Capuzzo-Dolcetta},
  {Mastrobuono-Battisti}, \& {Merritt}}]{antonini2012}
{Antonini}, F., {Capuzzo-Dolcetta}, R., {Mastrobuono-Battisti}, A., \&
  {Merritt}, D. 2012, \apj, 750, 111

\bibitem[{{Baganoff} {et~al.}(2001){Baganoff}, {Bautz}, {Brandt}, {Chartas},
  {Feigelson}, {Garmire}, {Maeda}, {Morris}, {Ricker}, {Townsley}, \&
  {Walter}}]{baganoff2001}
{Baganoff}, F.~K., {Bautz}, M.~W., {Brandt}, W.~N., {et~al.} 2001, \nat, 413,
  45

\bibitem[{{Baganoff} {et~al.}(2003){Baganoff}, {Maeda}, {Morris}, {Bautz},
  {Brandt}, {Cui}, {Doty}, {Feigelson}, {Garmire}, {Pravdo}, {Ricker}, \&
  {Townsley}}]{baganoff2003}
{Baganoff}, F.~K., {Maeda}, Y., {Morris}, M., {et~al.} 2003, \apj, 591, 891

\bibitem[{{Bahcall} \& {Wolf}(1976)}]{bahcall1976}
{Bahcall}, J.~N. \& {Wolf}, R.~A. 1976, \apj, 209, 214

\bibitem[{{Bahcall} \& {Wolf}(1977)}]{bahcall1977}
{Bahcall}, J.~N. \& {Wolf}, R.~A. 1977, \apj, 216, 883

\bibitem[{{Bartko} {et~al.}(2010){Bartko}, {Martins}, {Trippe}, {Fritz},
  {Genzel}, {Ott}, {Eisenhauer}, {Gillessen}, {Paumard}, {Alexander},
  {Dodds-Eden}, {Gerhard}, {Levin}, {Mascetti}, {Nayakshin}, {Perets},
  {Perrin}, {Pfuhl}, {Reid}, {Rouan}, {Zilka}, \& {Sternberg}}]{bartko2010}
{Bartko}, H., {Martins}, F., {Trippe}, S., {et~al.} 2010, \apj, 708, 834

\bibitem[{{Baumgardt} {et~al.}(2006){Baumgardt}, {Gualandris}, \& {Portegies
  Zwart}}]{baumgardt2006}
{Baumgardt}, H., {Gualandris}, A., \& {Portegies Zwart}, S. 2006, \mnras, 372,
  174

\bibitem[{{Blum} {et~al.}(1996){Blum}, {Sellgren}, \& {Depoy}}]{blum1996}
{Blum}, R.~D., {Sellgren}, K., \& {Depoy}, D.~L. 1996, \apj, 470, 864

\bibitem[{{Buchholz} {et~al.}(2009){Buchholz}, {Sch{\"o}del}, \&
  {Eckart}}]{buchholz2009}
{Buchholz}, R.~M., {Sch{\"o}del}, R., \& {Eckart}, A. 2009, \aap, 499, 483

\bibitem[{{Chandrasekhar}(1943)}]{chandrasekhar1943}
{Chandrasekhar}, S. 1943, \apj, 97, 255

\bibitem[{{Dale} {et~al.}(2009){Dale}, {Davies}, {Church}, \&
  {Freitag}}]{dale2009}
{Dale}, J.~E., {Davies}, M.~B., {Church}, R.~P., \& {Freitag}, M. 2009, \mnras,
  393, 1016

\bibitem[{{Do} {et~al.}(2009){Do}, {Ghez}, {Morris}, {Lu}, {Matthews}, {Yelda},
  \& {Larkin}}]{do2009stars}
{Do}, T., {Ghez}, A.~M., {Morris}, M.~R., {et~al.} 2009, \apj, 703, 1323

\bibitem[{{Dodds-Eden} {et~al.}(2011){Dodds-Eden}, {Gillessen}, {Fritz},
  {Eisenhauer}, {Trippe}, {Genzel}, {Ott}, {Bartko}, {Pfuhl}, {Bower},
  {Goldwurm}, {Porquet}, {Trap}, \& {Yusef-Zadeh}}]{dodds-eden2011}
{Dodds-Eden}, K., {Gillessen}, S., {Fritz}, T.~K., {et~al.} 2011, \apj, 728, 37

\bibitem[{{Duric}(2004)}]{duric2004}
{Duric}, N. 2004, {Advanced astrophysics}

\bibitem[{{Eckart} \& {Genzel}(1996)}]{eckart1996}
{Eckart}, A. \& {Genzel}, R. 1996, \nat, 383, 415

\bibitem[{{Eilon} {et~al.}(2009){Eilon}, {Kupi}, \& {Alexander}}]{Eilon2009}
{Eilon}, E., {Kupi}, G., \& {Alexander}, T. 2009, \apj, 698, 641

\bibitem[{{Eisenhauer} {et~al.}(2005){Eisenhauer}, {Genzel}, {Alexander},
  {Abuter}, {Paumard}, {Ott}, {Gilbert}, {Gillessen}, {Horrobin}, {Trippe},
  {Bonnet}, {Dumas}, {Hubin}, {Kaufer}, {Kissler-Patig}, {Monnet},
  {Str{\"o}bele}, {Szeifert}, {Eckart}, {Sch{\"o}del}, \&
  {Zucker}}]{eisenhauer2005}
{Eisenhauer}, F., {Genzel}, R., {Alexander}, T., {et~al.} 2005, \apj, 628, 246

\bibitem[{{Freitag}(2008)}]{freitag2008}
{Freitag}, M. 2008, in Astronomical Society of the Pacific Conference Series,
  Vol. 387, Massive Star Formation: Observations Confront Theory, ed.
  {H.~Beuther, H.~Linz, \& T.~Henning}, 247

\bibitem[{{Freitag} {et~al.}(2006){Freitag}, {Amaro-Seoane}, \&
  {Kalogera}}]{freitag2006}
{Freitag}, M., {Amaro-Seoane}, P., \& {Kalogera}, V. 2006, \apj, 649, 91

\bibitem[{{Fujii} {et~al.}(2009){Fujii}, {Iwasawa}, {Funato}, \&
  {Makino}}]{fujii2009}
{Fujii}, M., {Iwasawa}, M., {Funato}, Y., \& {Makino}, J. 2009, \apj, 695, 1421

\bibitem[{{Fujii} {et~al.}(2010){Fujii}, {Iwasawa}, {Funato}, \&
  {Makino}}]{fujii2010}
{Fujii}, M., {Iwasawa}, M., {Funato}, Y., \& {Makino}, J. 2010, \apjl, 716, L80

\bibitem[{{Genzel} {et~al.}(2000){Genzel}, {Pichon}, {Eckart}, {Gerhard}, \&
  {Ott}}]{genzel2000}
{Genzel}, R., {Pichon}, C., {Eckart}, A., {Gerhard}, O.~E., \& {Ott}, T. 2000,
  \mnras, 317, 348

\bibitem[{{Genzel} {et~al.}(2003){Genzel}, {Sch{\"o}del}, {Ott}, {Eisenhauer},
  {Hofmann}, {Lehnert}, {Eckart}, {Alexander}, {Sternberg}, {Lenzen},
  {Cl{\'e}net}, {Lacombe}, {Rouan}, {Renzini}, \&
  {Tacconi-Garman}}]{genzel2003stars}
{Genzel}, R., {Sch{\"o}del}, R., {Ott}, T., {et~al.} 2003, \apj, 594, 812

\bibitem[{{Ghez} {et~al.}(2003){Ghez}, {Duch{\^e}ne}, {Matthews}, {Hornstein},
  {Tanner}, {Larkin}, {Morris}, {Becklin}, {Salim}, {Kremenek}, {Thompson},
  {Soifer}, {Neugebauer}, \& {McLean}}]{ghez2003}
{Ghez}, A.~M., {Duch{\^e}ne}, G., {Matthews}, K., {et~al.} 2003, \apjl, 586,
  L127

\bibitem[{{Ghez} {et~al.}(1998){Ghez}, {Klein}, {Morris}, \&
  {Becklin}}]{ghez1998}
{Ghez}, A.~M., {Klein}, B.~L., {Morris}, M., \& {Becklin}, E.~E. 1998, \apj,
  509, 678

\bibitem[{{Gillessen} {et~al.}(2009{\natexlab{a}}){Gillessen}, {Eisenhauer},
  {Fritz}, {Bartko}, {Dodds-Eden}, {Pfuhl}, {Ott}, \&
  {Genzel}}]{gillessen2009s2}
{Gillessen}, S., {Eisenhauer}, F., {Fritz}, T.~K., {et~al.} 2009{\natexlab{a}},
  \apjl, 707, L114

\bibitem[{{Gillessen} {et~al.}(2009{\natexlab{b}}){Gillessen}, {Eisenhauer},
  {Trippe}, {Alexander}, {Genzel}, {Martins}, \& {Ott}}]{gillessen2009stars}
{Gillessen}, S., {Eisenhauer}, F., {Trippe}, S., {et~al.} 2009{\natexlab{b}},
  \apj, 692, 1075

\bibitem[{{Gould} \& {Quillen}(2003)}]{gould2003}
{Gould}, A. \& {Quillen}, A.~C. 2003, \apj, 592, 935

\bibitem[{{Gualandris} {et~al.}(2010){Gualandris}, {Gillessen}, \&
  {Merritt}}]{gualandris2010}
{Gualandris}, A., {Gillessen}, S., \& {Merritt}, D. 2010, \mnras, 409, 1146

\bibitem[{{Hansen} \& {Milosavljevi{\'c}}(2003)}]{hansen2003}
{Hansen}, B.~M.~S. \& {Milosavljevi{\'c}}, M. 2003, \apjl, 593, L77

\bibitem[{{Hopman} \& {Alexander}(2006{\natexlab{a}})}]{HA2006}
{Hopman}, C. \& {Alexander}, T. 2006{\natexlab{a}}, \apj, 645, 1152

\bibitem[{{Hopman} \& {Alexander}(2006{\natexlab{b}})}]{hopman2006}
{Hopman}, C. \& {Alexander}, T. 2006{\natexlab{b}}, \apjl, 645, L133

\bibitem[{{Kim} {et~al.}(2004){Kim}, {Figer}, \& {Morris}}]{kim2004}
{Kim}, S.~S., {Figer}, D.~F., \& {Morris}, M. 2004, \apjl, 607, L123

\bibitem[{{Kocsis} \& {Tremaine}(2011)}]{kocsis2011}
{Kocsis}, B. \& {Tremaine}, S. 2011, \mnras, 412, 187

\bibitem[{{Levin} {et~al.}(2005){Levin}, {Wu}, \& {Thommes}}]{levin2005}
{Levin}, Y., {Wu}, A., \& {Thommes}, E. 2005, \apj, 635, 341

\bibitem[{{Lightman} \& {Shapiro}(1977)}]{lightman1977}
{Lightman}, A.~P. \& {Shapiro}, S.~L. 1977, \apj, 211, 244

\bibitem[{{L{\"o}ckmann} {et~al.}(2009){L{\"o}ckmann}, {Baumgardt}, \&
  {Kroupa}}]{loeckmann2009}
{L{\"o}ckmann}, U., {Baumgardt}, H., \& {Kroupa}, P. 2009, \mnras, 398, 429

\bibitem[{{Madigan} {et~al.}(2009){Madigan}, {Levin}, \&
  {Hopman}}]{madigan2009}
{Madigan}, A.-M., {Levin}, Y., \& {Hopman}, C. 2009, \apjl, 697, L44

\bibitem[{{Ma{\'{\i}}z Apell{\'a}niz} \& {{\'U}beda}(2005)}]{maiz2005}
{Ma{\'{\i}}z Apell{\'a}niz}, J. \& {{\'U}beda}, L. 2005, \apj, 629, 873

\bibitem[{{Martins} {et~al.}(2008){Martins}, {Gillessen}, {Eisenhauer},
  {Genzel}, {Ott}, \& {Trippe}}]{martins2008}
{Martins}, F., {Gillessen}, S., {Eisenhauer}, F., {et~al.} 2008, \apjl, 672,
  L119

\bibitem[{{Merritt}(2010)}]{merritt2010}
{Merritt}, D. 2010, \apj, 718, 739

\bibitem[{{Merritt}(2012)}]{Merritt2012}
{Merritt}, D. 2012, {Black Holes and the Dynamics of Galactic Nuclei}
  (Princeton University Press)

\bibitem[{{Merritt} {et~al.}(2010){Merritt}, {Alexander}, {Mikkola}, \&
  {Will}}]{merritt2010RR}
{Merritt}, D., {Alexander}, T., {Mikkola}, S., \& {Will}, C.~M. 2010, \prd, 81,
  062002

\bibitem[{{Merritt} {et~al.}(2009){Merritt}, {Gualandris}, \&
  {Mikkola}}]{merritt2009}
{Merritt}, D., {Gualandris}, A., \& {Mikkola}, S. 2009, \apjl, 693, L35

\bibitem[{{Mouawad} {et~al.}(2005){Mouawad}, {Eckart}, {Pfalzner},
  {Sch{\"o}del}, {Moultaka}, \& {Spurzem}}]{mouawad2005}
{Mouawad}, N., {Eckart}, A., {Pfalzner}, S., {et~al.} 2005, Astronomische
  Nachrichten, 326, 83

\bibitem[{{Murphy} {et~al.}(1991){Murphy}, {Cohn}, \& {Durisen}}]{murphy1991}
{Murphy}, B.~W., {Cohn}, H.~N., \& {Durisen}, R.~H. 1991, \apj, 370, 60

\bibitem[{{Paumard} {et~al.}(2006){Paumard}, {Genzel}, {Martins}, {Nayakshin},
  {Beloborodov}, {Levin}, {Trippe}, {Eisenhauer}, {Ott}, {Gillessen}, {Abuter},
  {Cuadra}, {Alexander}, \& {Sternberg}}]{paumard2006}
{Paumard}, T., {Genzel}, R., {Martins}, F., {et~al.} 2006, \apj, 643, 1011

\bibitem[{{Perets} \& {Gualandris}(2010)}]{perets2010}
{Perets}, H.~B. \& {Gualandris}, A. 2010, \apj, 719, 220

\bibitem[{{Perets} {et~al.}(2009){Perets}, {Gualandris}, {Kupi}, {Merritt}, \&
  {Alexander}}]{perets2009}
{Perets}, H.~B., {Gualandris}, A., {Kupi}, G., {Merritt}, D., \& {Alexander},
  T. 2009, \apj, 702, 884

\bibitem[{{Perets} {et~al.}(2007){Perets}, {Hopman}, \&
  {Alexander}}]{perets2007}
{Perets}, H.~B., {Hopman}, C., \& {Alexander}, T. 2007, \apj, 656, 709

\bibitem[{{Preto} \& {Amaro-Seoane}(2010)}]{preto2010}
{Preto}, M. \& {Amaro-Seoane}, P. 2010, \apjl, 708, L42

\bibitem[{{Rauch} \& {Tremaine}(1996)}]{RT96}
{Rauch}, K.~P. \& {Tremaine}, S. 1996, \na, 1, 149

\bibitem[{{Rubilar} \& {Eckart}(2001)}]{rubilar2001}
{Rubilar}, G.~F. \& {Eckart}, A. 2001, \aap, 374, 95

\bibitem[{{Sabha} {et~al.}(2010){Sabha}, {Witzel}, {Eckart}, {Buchholz},
  {Bremer}, {Gie{\ss}{\"u}bel}, {Garc{\'{\i}}a-Mar{\'{\i}}n}, {Kunneriath},
  {Muzic}, {Sch{\"o}del}, {Straubmeier}, {Zamaninasab}, \&
  {Zernickel}}]{sabha2010}
{Sabha}, N., {Witzel}, G., {Eckart}, A., {et~al.} 2010, \aap, 512, A2

\bibitem[{{Sabha} {et~al.}(2011){Sabha}, {Witzel}, {Eckart}, {Zamaninasab},
  {Sch{\"o}del}, {Zernickel}, {Garc{\'{\i}}a-Mar{\'{\i}}n}, {Kunneriath},
  {Moultaka}, {Mu{\v z}i{\'c}}, \& {Straubmeier}}]{sabha2011}
{Sabha}, N., {Witzel}, G., {Eckart}, A., {et~al.} 2011, in Astronomical Society
  of the Pacific Conference Series, Vol. 439, Astronomical Society of the
  Pacific Conference Series, ed. {M.~R.~Morris, Q.~D.~Wang, \& F.~Yuan}, 313

\bibitem[{{Salaris} \& {Cassisi}(2005)}]{salaris2005}
{Salaris}, M. \& {Cassisi}, S. 2005, {Evolution of Stars and Stellar
  Populations}

\bibitem[{{Sazonov} {et~al.}(2011){Sazonov}, {Sunyaev}, \&
  {Revnivtsev}}]{sazonov2011}
{Sazonov}, S., {Sunyaev}, R., \& {Revnivtsev}, M. 2011, \mnras, 1987

\bibitem[{{Sch{\"o}del} {et~al.}(2007){Sch{\"o}del}, {Eckart}, {Alexander},
  {Merritt}, {Genzel}, {Sternberg}, {Meyer}, {Kul}, {Moultaka}, {Ott}, \&
  {Straubmeier}}]{schoedel2007}
{Sch{\"o}del}, R., {Eckart}, A., {Alexander}, T., {et~al.} 2007, \aap, 469, 125

\bibitem[{{Sch{\"o}del} {et~al.}(2010){Sch{\"o}del}, {Najarro}, {Muzic}, \&
  {Eckart}}]{schoedel2010}
{Sch{\"o}del}, R., {Najarro}, F., {Muzic}, K., \& {Eckart}, A. 2010, \aap, 511,
  A18

\bibitem[{{Sch{\"o}del} {et~al.}(2002){Sch{\"o}del}, {Ott}, {Genzel},
  {Hofmann}, {Lehnert}, {Eckart}, {Mouawad}, {Alexander}, {Reid}, {Lenzen},
  {Hartung}, {Lacombe}, {Rouan}, {Gendron}, {Rousset}, {Lagrange}, {Brandner},
  {Ageorges}, {Lidman}, {Moorwood}, {Spyromilio}, {Hubin}, \&
  {Menten}}]{schoedel2002}
{Sch{\"o}del}, R., {Ott}, T., {Genzel}, R., {et~al.} 2002, \nat, 419, 694

\bibitem[{{Timmes} {et~al.}(1996){Timmes}, {Woosley}, \& {Weaver}}]{timmes1996}
{Timmes}, F.~X., {Woosley}, S.~E., \& {Weaver}, T.~A. 1996, \apj, 457, 834

\bibitem[{{Witzel} {et~al.}(2012){Witzel}, {Eckart}, {Bremer}, {Zamaninasab},
  {Shahzamanian}, {Valencia-S.}, {Sch{\"o}del}, {Karas}, {Lenzen}, {Marchili},
  {Sabha}, {Garc{\'{\i}}a-Mar{\'{\i}}n}, {Buchholz}, {Kunneriath}, \&
  {Straubmeier}}]{witzel2012}
{Witzel}, G., {Eckart}, A., {Bremer}, M., {et~al.} 2012, \apj, submitted

\bibitem[{{Yusef-Zadeh} {et~al.}(2012){Yusef-Zadeh}, {Bushouse}, \&
  {Wardle}}]{yusefzadeh2012}
{Yusef-Zadeh}, F., {Bushouse}, H., \& {Wardle}, M. 2012, \apj, 744, 24

\bibitem[{{Zucker} {et~al.}(2006){Zucker}, {Alexander}, {Gillessen},
  {Eisenhauer}, \& {Genzel}}]{zucker2006}
{Zucker}, S., {Alexander}, T., {Gillessen}, S., {Eisenhauer}, F., \& {Genzel},
  R. 2006, \apjl, 639, L21

\end{thebibliography}

\end{document}